\documentclass[aps,prb,twocolumn,showpacs,apsrev]{revtex4-1}
\usepackage[T1]{fontenc}
\usepackage{lmodern}
\usepackage[english]{babel}
\usepackage{amsmath}
\usepackage{amsfonts}
\usepackage{graphicx} 
\usepackage{comment}
\usepackage{xfrac}
\usepackage{color}
\usepackage{multirow}
\usepackage{hyperref}
\usepackage{braket}
\usepackage{pstricks}
\usepackage{pgfplots}
\pgfplotsset{compat=newest}

\hypersetup{
	colorlinks = true,
	pdftitle = {},
	pdfauthor = {},
	pdfkeywords = {},
	linkcolor = blue,
	citecolor = blue,
	filecolor = black,
	urlcolor = magenta
}

\newcommand{\dd}{\mathrm{d}}
\newcommand{\ee}{\mathrm{e}}
\newcommand{\img}{\mathrm{i}}
\newcommand{\tcm}{T_\text{c}^{M}}
\newcommand{\tcp}{T_\text{c}^{P}}

\newcommand{\tcC}{T_\text{c}^{C}}
\newcommand{\smo}{SrMnO$_3$}

\graphicspath{{./graphics/}}

\begin{document}

\title{Prediction of a Giant Magnetoelectric Cross-Caloric Effect Around a Tetracritical Point in Multiferroic SrMnO$_3$}
 
\author{Alexander Edstr\"om}
\author{Claude Ederer}
\affiliation{Materials Theory, ETH Z\"urich, Wolfgang-Pauli-Strasse 27, 8093 Z\"urich, Switzerland}

%\date{\today}

\begin{abstract}
We study the magnetoelectric and electrocaloric response of strain-engineered, multiferroic \smo, using a phenomenological Landau theory with all parameters obtained from \emph{first-principles}-based calculations. This allows to make realistic and materials-specific predictions about the magnitude of the corresponding effects. 
We find that in the vicinity of a tetracritical point, where magnetic and ferroelectric phase boundaries intersect, an electric field has a huge effect on the antiferromagnetic order, corresponding to a magnetoelectric response several orders of magnitude larger than in conventional linear magnetoelectrics.
Furthermore, the strong magnetoelectric coupling leads to a magnetic, cross-caloric contribution to the electrocaloric effect, which increases the overall caloric response by about 60\%.
This  opens up new potential applications of antiferromagnetic multiferroics in the context of environmentally friendly solid state cooling technologies. 
\end{abstract}

\maketitle

%\section{Introduction}
\emph{Introduction - }
Caloric effects in ferroic materials, where application/removal of external fields (magnetic, electric, or stress) can result in significant temperature changes, potentially allow for the development of clean and energy-efficient cooling technologies~\cite{Faehler_et_al:2011,Moya2014}.
More recently, there has been growing interest in so-called multicaloric effects~\cite{stern-taulats2018,VOPSON20122067}, where more than one type of caloric effect can occur simultaneously, possibly allowing to further optimize the total caloric response. The thermodynamic theory of multicaloric effects has been discussed in some detail~\cite{MENG2013567,Anand_2014,planes_multical}.
However, most specific studies have been focusing on combining either electrocaloric or magnetocaloric with elastocaloric effects, thereby using applied stress or strain as an additional control parameter to enhance the overall caloric response~\cite{Lisenkov_et_al:2013,PhysRevB.94.214113,doi:10.1002/adma.201404725} and/or to reduce irreversibility problems~\cite{Liu2012,PhysRevB.91.224421,Liu2016,Gottschall2018}.
Multicaloric effects in (single phase) materials combining magnetic and ferroelectric (FE) order, meanwhile, have remained relatively unexplored~\cite{Moya2014,stern-taulats2018}, perhaps due to challenges in finding suitable materials.

Multiferroic materials with coexisting magnetic and FE orders have received much attention, not only because of a broad fundamental interest, but also due to promises of technological applications~\cite{Spaldin/Cheong/Ramesh:2010,Spaldin/Ramesh:2019}.
Often, however, their practical usefulness is hindered by low ordering temperatures or weak magnetoelectric (ME) coupling. Additionally, most magnetic ferroelectrics are in fact antiferromagnetic (AFM), which restricts their potential applications, since the AFM order does not couple to a homogeneous magnetic field.
Here we show that an AFM multiferroic can, nevertheless, exhibit a very strong cross-caloric magnetic contribution to the electrocaloric effect (ECE)~\cite{multical_note}. 

Since caloric effects are generally largest near the relevant phase transitions, a strong cross-caloric effect can be expected near a so-called \emph{tetracritical point} (TCP)~\cite{LL_statphys}, where the critical temperatures of the two phase transitions coincide. Such a TCP has recently been predicted to occur in strained SrMnO$_3$~\cite{PhysRevMaterials.2.104409}; its existence can also be inferred from previous theoretical~\cite{PhysRevLett.104.207204} and experimental~\cite{Becher2015,acs.nanolett.5b04455,PhysRevB.97.235135} work.
While perovskite structure bulk SrMnO$_3$ is a cubic paraelectric G-type antiferromagnet~\cite{JPSJ.37.275}, it develops a FE distortion under tensile epitaxial strain~\cite{PhysRevLett.104.207204,Becher2015,acs.nanolett.5b04455,PhysRevB.97.235135}. Thereby, the FE critical temperature increases strongly with strain~\cite{PhysRevMaterials.2.104409}, while the AFM N\'eel temperature is less affected, resulting in an intersection of the FE and AFM phase boundaries at a certain strain value, and thus a TCP.
Furthermore, since the Mn cation carries the magnetic moment and also takes part in the FE distortion, \smo{} is expected to exhibit strong ME coupling, which is also implied by recent studies reporting a particularly strong spin-phonon coupling in this material~\cite{PhysRevB.84.104440,PhysRevB.89.064308}. 

In this work, we explore ME coupling effects and cross-caloric response in \smo{} by constructing a Landau-type theoretical model considering all relevant magnetic and ferroelectric order parameters. We extract all parameters entering the free energy from {\emph{first principles}}-based calculations, thus allowing for a realistic materials-specific description. We then apply the model to study ME coupling phenomena around the multiferroic TCP in \smo. We show that an applied electric field has a strong effect on the AFM order, shifting its critical temperature and increasing the corresponding order parameter, thereby drastically changing the entropy of the magnetic sub-system. This results in a huge magnetic cross-caloric contribution to the ECE, which is increased by about 60\,\% due to the ME coupling.

%\section{Methods}\label{sec.meth}
\emph{Methods -}
SrMnO$_3$ under epitaxial strain is predicted to show a number of different magnetic phases, including G, C and A-type AFM~\cite{Wollan/Koehler:1955}, and possibly also ferromagnetic (FM) at large strains (near 5\%)~\cite{PhysRevLett.104.207204,PhysRevMaterials.2.104409}. Furthermore, in the cubic structure, there are three different degenerate $\mathbf{q}$-vectors corresponding to each of the A-type and C-type AFM orders. When the cubic symmetry is broken this degeneracy is also broken. Thus, we consider eight magnetic order parameters: FM [$\mathbf{q}=(0,0,0)$], G [$\mathbf{q}=(1,1,1)$], A [$\mathbf{q}=(0,0,1)$ or $\mathbf{q}=(0,1,0)$ or $\mathbf{q}=(1,0,0)$] and C [$\mathbf{q}=(1,1,0)$ or $\mathbf{q}=(1,0,1)$ or $\mathbf{q}=(0,1,1)$], where the reciprocal space vectors are given in units of $\pi$ divided by the real space lattice constant along that direction. This includes all magnetic orders that have been reported to appear in \smo~\cite{PhysRevMaterials.2.104409}. Each of these magnetic order parameters can couple to the polar order $P$ that emerges under strain. Hence, we consider a Landau free energy of the form 
\begin{widetext}
\begin{equation}\label{eq.LandauF}
	\mathcal{F}_q =   \frac{1}{2} a_P(T,\eta) P^2 + \frac{b_P}{4} P^4 + \frac{1}{2} a_q(T,\eta) M_q^2 + \frac{b_q}{4} M_q^4  + \frac{\lambda_q(\eta)}{2} M_q^2P^2  - EP ,
\end{equation}
\end{widetext}
for each magnetic order parameter $M_\mathbf{q} = \frac{1}{N} \sum_i^N	\ee^{\img \mathbf{q} \cdot \mathbf{R}_i} \langle S_i \rangle$, where $\langle S_i \rangle$ is the thermodynamic average of the normalized spin at site $\mathbf{R}_i$, projected on the spin-quantization axis, and $N$ is the number of spins. The strain and temperature dependence enters in the second order coefficients as $a_P = \alpha_P(T-T_{0}^{P}) + c_P \eta $ and $ a_q = \alpha_q(T-T_{0}^{q}) + c_q \eta$. 
At each strain $\eta$, temperature $T$, and electric field $E$, the free energy $\mathcal{F}_q$ is minimized with respect to $P$ and $M_q$, and the free energy is determined from ${\cal F} = \min_q {\cal F}_q$. The $q$ which corresponds to the lowest free energy defines the equilibrium magnetic phase at that point in the phase diagram. 

All parameters in Eq.~\eqref{eq.LandauF} were determined from total energy calculations using density functional theory (DFT) and DFT-based effective Hamiltonian simulations~\cite{PhysRevMaterials.2.104409,suppl}. Specifically, the magnetic parameters were obtained by mapping DFT total energy calculations on a Heisenberg Hamiltonian and extracting exchange interaction parameters. By calculating exchange interactions as functions of strain, the coupling between strain and magnetism was obtained, while exchange interactions evaluated with FE structural distortions allowed the determination of the biquadratic magnetoelectric coupling coefficients $\lambda_q$. 
The purely ferroelectric parameters are determined from the strain-dependent transition temperature and saturation polarization obtained from first-principles-based effective Hamiltonians~\cite{PhysRevMaterials.2.104409}, and from DFT calculated elastic/electro-strictive parameters.

\begin{figure}[bt]
	\centering
	\includegraphics[trim={0cm 0 0 0},clip,width=0.495\textwidth]{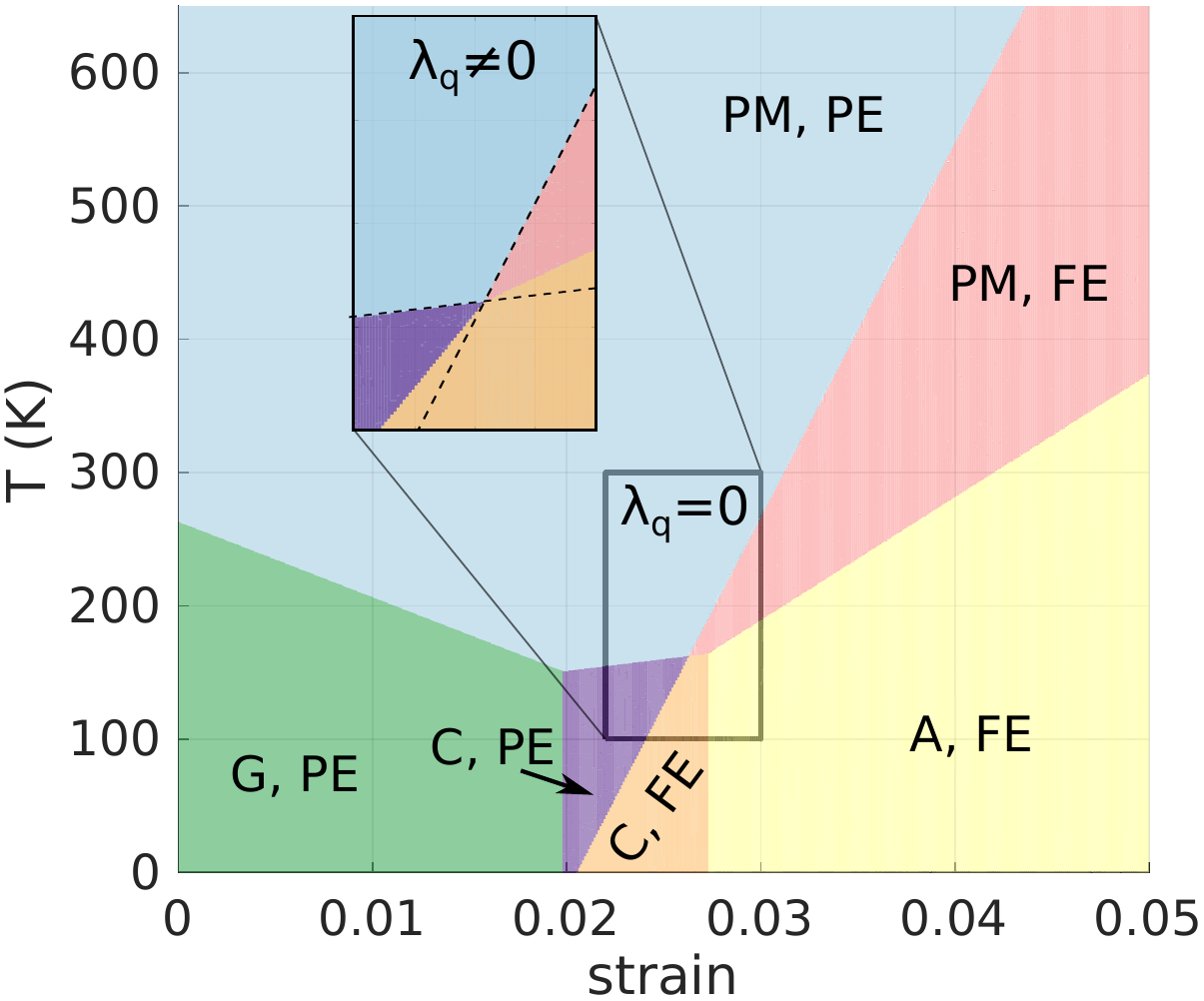}
	\caption{Ferroic phase diagram of SrMnO$_3$ at zero applied field obtained within our Landau theory for the case without ME coupling ($\lambda_q=0$). The inset shows the effect of non-zero ME coupling on the region around the TCP. The dashed lines in the inset indicate the FE-PE and the C-paramagnetic (PM) phase boundaries with $\lambda_q=0$.}
	\label{fig.phasediag}
\end{figure}

%\section{Results}
\emph{Results -}
We first consider the case without ME coupling and zero applied field, i.e., $\lambda_q=0$ and $E=0$, and minimize the free energy in Eq.~\eqref{eq.LandauF} with respect to the various order parameters for different temperatures and strains in the range $0 \leq \eta \leq 5\%$. Identifying the phases with the lowest free energy for each strain and temperature results in the phase diagram shown in Fig.~\ref{fig.phasediag}, which agrees well with the one from our previous study using microscopic first-principles-based Hamiltonians~\cite{PhysRevMaterials.2.104409}.  For small $T$ and $\eta$, there is a G-type AFM paraelectric (PE) phase, while at approximately 2\% strain there is a transition into a FE region and also a change to C-type [$\mathbf{q}=(1,0,1)$] AFM order. For large strain and low temperatures, an A-type [$\mathbf{q}=(0,0,1)$] AFM FE region appears. In the following C and A-type AFM always refer to $\mathbf{q}$-vectors $(1,0,1)$ and $(0,0,1)$, since these are the only ones that appear in the phase diagram.
We note that the ferromagnetism that has been predicted for large strains is only stabilized due to its coupling to the FE order~\cite{PhysRevMaterials.2.104409}, which at this point is not yet included in our free energy. Most notably, the phase diagram in Fig.~\ref{fig.phasediag} reveals a TCP where the magnetic and FE critical temperatures coincide within the region with C-type AFM order at $\eta_\text{tcp}=2.63\%$ and $T_\text{tcp}=162~\mathrm{K}$.

Next, we evaluate the strain-dependent ME coupling parameters, $\lambda_q$, by computing magnetic exchange interactions as functions of the FE displacements for different strains. 
As shown in the supplementary material~\cite{suppl}, it turns out that the lowest order biquadratic coupling in Eq.~\eqref{eq.LandauF} is insufficient to describe the variation of the exchange couplings for large polarization, which occurs in the region of the phase diagram with large strain and low temperatures. 
A satisfactory description of this region would require coupling terms of higher order in $P$, which, however, would require additional higher order terms to guarantee stable, physical solutions, and thus more parameters in the free energy.
In the following, we therefore focus on the part of the phase diagram which is most interesting in the present context, i.e., the region around the TCP, where both order parameters are small~\cite{ME-coupling-note}. 

For the C-type order relevant around the TCP, we find a negative ME coupling, which varies relatively weakly with strain. We point out that, previously, a positive ME coupling coefficient $\lambda_\mathrm{G}$ has been found for cubic 
Sr$_{1-x}$Ba$_x$MnO$_3$ ~\cite{PhysRevLett.107.137601,PhysRevLett.109.107601}, meaning that G-type AFM order and ferroelectricity couple unfavorably. This is indeed consistent with our results~\cite{suppl}. However, we also find that the coupling coefficients differ for different types of magnetic order and, furthermore, are strongly strain-dependent. 

\begin{figure}[hbt]
	\centering
	\includegraphics[trim={0cm 0 0 0},clip,width=0.495\textwidth]{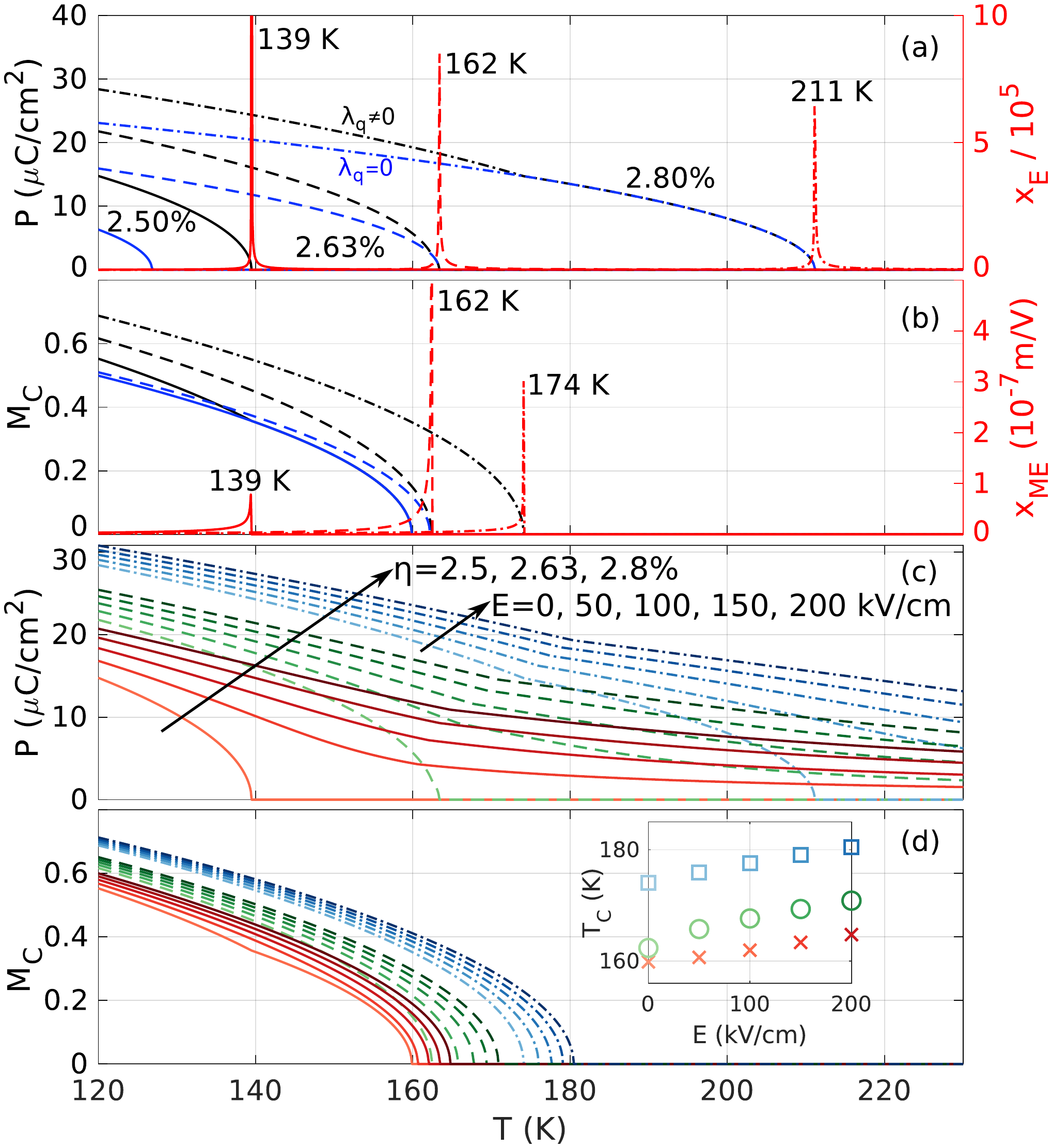}
	\caption{(a) and (b): Order parameters (black, left) and susceptibilities (red, right) as functions of temperature for the three strains of 2.5\% (solid line), 2.63\% (dashed line) and 2.8\% (dashed dotted line). (a) shows the electric polarization and electric susceptibility, while (b) shows the magnetic order parameter and ME susceptibility. The order parameters for zero ME coupling are shown in blue. (c) and (d): Temperature dependence of FE (c) and magnetic (d) order parameters for strains of 2.5\% (red solid lines), 2.63\% (green dashed lines) and 2.8\% (blue dashed dotted lines) with applied electric fields of 0, 50, 100, 150 and 200 kV/cm. The darker colors correspond to larger fields. The inset in (d) shows the magnetic transition temperature as function of electric field, with color coding corresponding to the main plot.}
	\label{fig.OrdPar_T}
\end{figure}

The zero field phase diagram for the region $2.2\% \leq \eta \leq 3.0\%$ and $100~\mathrm{K} \leq T \leq 300~\mathrm{K}$, now including ME coupling, is shown in the inset of Fig.~\ref{fig.phasediag}. One drastic effect of the coupling is that it eliminates the A-type AFM region from the phase diagram. This is because $\lambda_A$ is found to be strongly positive and since A-type order only appears in the FE region, it is highly unfavored by the coupling, while C-type is favored. 
In contrast, the coupling does not alter the position of the TCP, since both $M_C$ and $P$, and thus the effect of the coupling term, vanish at this point. 
Away from the TCP, the upper of the two ordering temperatures also remains unaltered, while the lower one is increased by the negative ME coupling. 
This can also be seen from Figs.~\ref{fig.OrdPar_T}(a) and (b), which show the temperature dependence of the FE polarization $P$ and the C-AFM order parameter $M_\text{C}$, both with (black) and without (blue) ME coupling, for three different strain values. 
At $\eta = 2.80\%$ (where $\tcC<\tcp$), $\tcp$ is unaffected, while the magnetic order is changed from A-type to C-type with an increase in ordering temperature from $T_c^A = 170~\mathrm{K}$ to $\tcC = 174~\mathrm{K}$. 
In addition, the polarization is unaffected by the coupling at temperatures above $\tcC=174~\text{K}$, while the coupling enhances the polarization at lower temperatures, producing a kink in $P(T)$ at $\tcC$. 
The analogous behavior, but with the roles of $P$ and $M_\text{C}$ exchanged, is observed at $\eta = 2.50\%$ (where $\tcC>\tcp$).
Here, the coupling does not alter $\tcm$, while it shifts $\tcp$ from 127\,K to 139\,K, resulting in a kink in $M_\text{C}(T)$ at $\tcp = 139~\text{K}$.
On the other hand, at $\eta_\text{tcp} = 2.63\%$, the coinciding critical temperatures are unaltered by the coupling term. However, below $T_\text{tcp}=162~\text{K}$, both order parameters are enhanced compared to the case with $\lambda_q=0$. 
This behavior is consistent with the general phenomenological theory outlined in Ref.~[\onlinecite{planes_multical}], where it was also shown that both transitions remain second order if $\lambda_q^2 < b_q b_p$ (or if $\lambda_q > 0$).  According to our results this condition is fulfilled for every magnetic order and strain considered. 

The zero-field electric susceptibility $\chi_E=\frac{\dd P}{\dd E}|_{E=0}$ (for the case with ME coupling), is also plotted in Fig.~\ref{fig.OrdPar_T}(a) (red, right $y$-axis). As expected, this susceptibility diverges at the FE transitions. Additionally, the magnetoelectric susceptibility 
\begin{equation}
\chi_{ME} = \left. \frac{\dd M_q}{\dd E} \right|_{E=0} =
	\begin{cases}
      0, & \text{if}\ M_q=0~\text{or}~P=0 \\
      -\frac{\lambda_q P}{b_q M_q} \chi_E, & \text{if}\ M_q\neq0~\text{and}~P\neq0
    \end{cases} 
\end{equation}
is plotted in Fig.~\ref{fig.OrdPar_T}(b) (red, right $y$-axis). This quantity describes the magnetic response to an applied electric field and is non-zero only in the multiferroic regions of the phase diagram, i.e., where both magnetic and FE order parameters are non-zero.
The ME susceptibility then diverges at the lower of the two transition temperatures, either because $\chi_E$ diverges if the FE transition is lower, or because $M_q \rightarrow 0$ if the magnetic transition is lower. Thus, $\chi_{ME}$ diverges at $\tcp = 139~\text{K}$ for $\eta = 2.50\%$, at $\tcm = \tcp = 162~\text{K}$ for $\eta_\mathrm{tcp} = 2.63\%$, and at $\tcm = 174~\text{K}$ for $\eta = 2.80\%$. The divergence is particularly pronounced at $\eta_\mathrm{tcp}$, where $\chi_E$ diverges simultaneously as $M_q \rightarrow 0$, causing $\chi_\mathrm{ME}$ to diverge as $(T_\mathrm{c} - T)^{-1}$ instead of $(T_\mathrm{c} - T)^{-1/2}$ when the relevant critical temperature $T_\mathrm{c}$ is approached from below~\cite{critexpcomment}. 

We now discuss the effect of applying a finite electric field. In Fig.~\ref{fig.OrdPar_T}(c)-(d), the FE and magnetic order parameters are plotted as functions of temperature for the previously discussed strain values and various applied electric fields. As expected, an electric field induces a finite electric polarization at all temperatures, which however, decreases towards high $T$, and thus removes the second order FE transition.
The effect on the magnetic order parameter is markedly different. While the electric field enhances also $M_\text{C}$, due to the negative sign of $\lambda_\text{C}$, the magnetic order parameter still shows a second order transition, and is identically zero above the corresponding transition temperature. The magnetic transition temperature is, however, field dependent and the inset of Fig.~\ref{fig.OrdPar_T}(d) shows $\tcC$ as a function of applied electric field. The increase in $\tcC$ with $E$ appears close to linear, and an applied field of 100 keV/cm$^2$ increases $\tcC$ by 2.1 K for $\eta=2.5\%$, by 5.3 K for $\eta=2.63\%$, and by 3.5 K for $\eta=2.8\%$. The largest effect of the electric field on $\tcC$ is thus found at $\eta_\mathrm{tcp}$.

We note that SrMnO$_3$ is not a linear ME material. Nevertheless, in order to get a better idea of the magnitude of the electric field effect on $M_\text{C}$, one can see from Fig.~\ref{fig.OrdPar_T}(d) that an electric field of $50~\mathrm{kV/cm}$ alters $M_\mathrm{C}$ by about 0.15 at the TCP. Considering a Mn magnetic moment of $3\mu_\mathrm{B}$, one can estimate an effective ME coefficient of $\alpha_\mathrm{eff} = \frac{\Delta M}{\Delta E} = 15 \cdot 10^{-3}~\mathrm{\Omega^{-1}}$, which is four orders of magnitude larger than that found in conventional linear magnetoelectrics such as Cr$_2$O$_3$~\cite{doi:10.1080/00150199408245099,PhysRevLett.101.117201}. 

Based on the electric field response of both FE and magnetic order parameters, we can now address the ECE in SrMnO$_3$.
From the results presented so far, it is apparent that, due to the negative ME coupling coefficient, an applied electric field has an ordering tendency on both the FE and magnetic subsystems, and hence reduces the entropy in both. This will result in a magnetic contribution to the ECE, referred to as cross-caloric~\cite{planes_multical}. 
The caloric response is quantified by the isothermal entropy change under field application or removal. From the free energy in Eq.~\eqref{eq.LandauF}, the entropy at a given temperature and field $E$ is $S(T,E)=-\left( \frac{\partial \mathcal{F} }{\partial T} \right)_E$, while the entropy change when increasing the field from 0 to $E$ is $\Delta S (T,E) = S(T,E) - S(T,0) = -\frac{1}{2} \alpha_P \left( P^2(T,E) - P^2(T,0)\right) -\frac{1}{2} \alpha_q \left( M_q^2(T,E) - M_q^2(T,0)\right)$. Here, the first term is the usual ECE, while the second term is the magnetic contribution, i.e., the cross-caloric response.

\begin{figure}[tbh]
	\centering
	\includegraphics[trim={0cm 0 0 0},clip,width=0.495\textwidth]{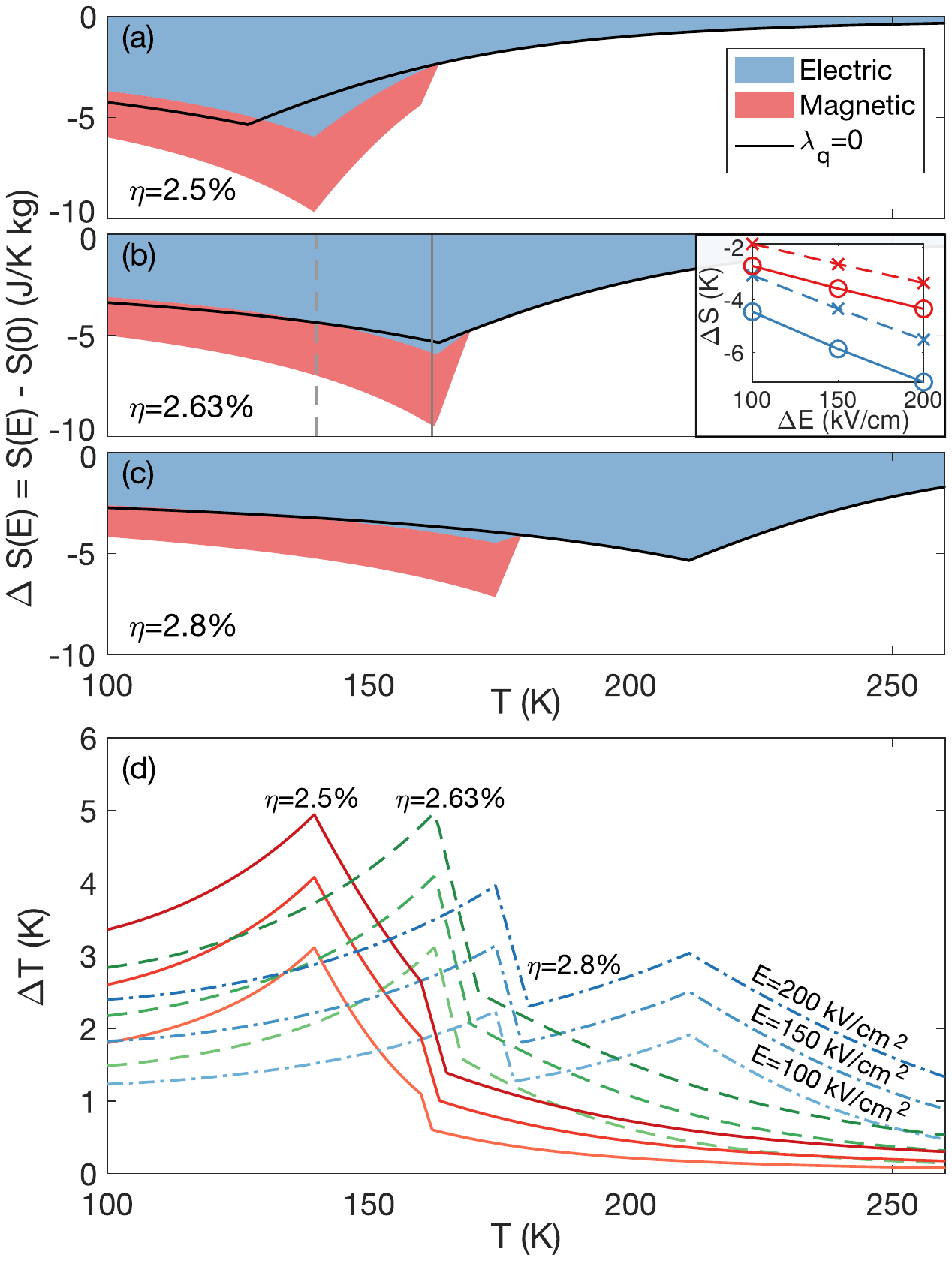}
	\caption{The ECE as function of temperature. (a)-(c) show the isothermal entropy change, as a field of 150~kV/cm is applied, for the three different strains of 2.5\%, 2.63\% and 2.8\%, respectively, while (b) also contains an inset showing the field dependence at $T=140~\mathrm{K}$ (dashed) and $T=162~\mathrm{K}$ (solid). The total entropy change is decomposed into magnetic (red) and electric (blue) contributions. Additionally, the result occurring without ME coupling ($\lambda_q=0$) is shown (black line). (d) shows an estimate of the adiabatic temperature change corresponding to the total isothermal entropy change for strains of 2.5\%, 2.63\% and 2.8\%, and applied electric fields of 100, 150 or 200 kV/cm. }
	\label{fig.Calorics}
\end{figure}

Fig.~\ref{fig.Calorics}(a)-(c) show the isothermal entropy change in SrMnO$_3$ as function of temperature for an applied field of 150\,kV/cm, at the three representative strain values discussed previously ($\eta=2.5\%$, $\eta=2.63\%$, and $\eta=2.8\%$). The total entropy change has been decomposed in magnetic and electric contributions and the ECE obtained without ME coupling ($\lambda_q = 0$) is also plotted as a black line.
The total caloric response exhibits features (peaks and/or kinks) at all critical temperatures (with or without field).
Generally, the electric contribution is non-zero at all temperatures and peaks at the zero field $\tcp{}$. Below but near $\tcC(E)$ it is enhanced compared to the case without ME coupling. For $\eta = 2.9\,\%$ this even leads to an additional small peak  at $\tcC(0)$.
Hence, the ME coupling can enhance the ECE not only by adding the magnetic cross-caloric effect, but also by enhancing the electric part.
The magnetic contribution vanishes above $\tcC(E)$, but rises sharply between $\tcC(E)$ and $\tcC(0)$, peaking at $\tcC(0)$, then slowly decreases again towards lower $T$, except for the case of $\eta=2.5\,\%$, where it actually peaks at $\tcp{}$.
This is related to the kink in $M_\text{C}(T)$ at this temperature for zero field (see Fig.~\ref{fig.OrdPar_T}(b)).
The inset in Fig.~\ref{fig.Calorics}(b) shows the magnetic and electric contributions to the entropy change at $\eta=2.63\%$ and temperatures 140~K and 162~K, as functions of applied electric field, which illustrates an approximately linear increase in the magnitude of the entropy change with the field. 

Most strikingly, at all three strains, the magnetic contribution reaches approximately 60\% of the electric contribution, or more than a third of the total entropy change. This is a result of particular relevance, since it shows that the ME cross-caloric effect can significantly increase the caloric response suitable for solid state cooling. Furthermore, the effect is of similar size for the three different strains, indicating that a very careful tuning of the two critical temperatures to coincide is not necessary.
It is also interesting to note that, in the case of $\eta = 2.8\%$, the largest total ECE is not obtained at the FE phase transition, but at the magnetic one. This is because it is the lower temperature phase transition in this case and thus the two contributions add up, while at the FE transition the magnetic contribution vanishes. 

Another instructive quantity to characterize caloric effects is the adiabatic temperature change $\Delta T$, which can be estimated from the entropy change $\Delta S$ via the thermodynamic relation $\dd T = -\frac{T}{C} \dd S$, where $C$ is the specific heat at constant field, without the contributions of the FE or magnetic degrees of freedom~\cite{Gruenebohm_et_al:2018}.
We use this relation to estimate $\Delta T \approx - \frac{T}{C} \Delta S$, assuming that $\Delta T \ll T$ and that $C$ varies negligibly over $\left[ T, T+\Delta T\right]$. For $C$, we use the temperature dependent phonon specific heat, which we obtained for cubic \smo, from frozen phonon calculations in the harmonic approximation~\cite{TOGO20151}. This results in a double counting of the phonon modes responsible for ferroelectricity, which might slightly underestimate $\Delta T$. The resulting $\Delta T$ is plotted in Fig.~\ref{fig.Calorics}(d), for the same strains as in (a)-(c), and three different applied fields. The largest temperature changes, for $E=200~\mathrm{kV/cm}$, are about 5 K. This is of the order of magnitude needed to be technologically relevant and of similar size as the ECE found in high performing electrocaloric materials for similar field strengths~\cite{Moya2014}. Although being estimates, the temperature changes in Fig.~\ref{fig.Calorics}(d) show that multiferroic perovskite oxides can indeed be of potential technological relevance within the area of solid state cooling. 

%\section{Summary and Conclusions}\label{sec.concl}
\emph{Summary and conclusions -}
We have used a Landau theory, allowing several magnetic order parameters to couple to a FE polarization, to study ME coupling phenomena around the TCP appearing in the strain-temperature phase diagram of SrMnO$_3$. Since all parameters entering the theory have been determined from {\emph{first principles}} DFT-based calculations, realistic materials specific predictions can be made without experimental input. The ME coupling is found to be enhanced at the TCP and a huge response to electric fields is observed in the magnetic order parameter. Investigating the ECE, we find a large cross-caloric contribution due to the electric-field-induced magnetic entropy change, resulting in an increase of about 60\% in the total caloric response. This provides a new way for greatly enhancing caloric effects for solid state cooling applications, by using multiferroic materials with coupled magnetic and electric order parameters. It also provides a unique example where AFM order in a multiferroic material can be of great practical usefulness. Recent work proving that highly strained multiferroic films of SrMnO$_3$ can be grown~\cite{PhysRevB.97.235135} is promising regarding the experimental verification of these results, while similar studies on Ba-doped systems\cite{PhysRevLett.107.137601,PhysRevMaterials.2.054408,doi:10.1063/1.5090824} would also be of interest. Further insights could also be obtained by studies using other computational methods, e.g., based on microscopic models for coupled spin-lattice dynamics~\cite{PhysRevLett.99.227602,PhysRevB.99.104302}. 

%\section{Acknowledgments}
\emph{Acknowledgments -}
A.E. is grateful to Quintin Meier for discussions. This work was supported by the Swiss National Science Foundation (project code 200021E-162297) and the German Science Foundation under the priority program SPP 1599 (``Ferroic Cooling''). Computational work was performed on resources provided by the Swiss National Supercomputing Centre (CSCS).  

%\appendix
%\section{Appendix title}\label{AppA}

\bibliography{literature}{}

\begin{thebibliography}{40}
\bibitem [{mul()}]{multical_note}%
  \BibitemOpen
  \href@noop {} {}\bibinfo {note} {We use the terminology of
  Ref.~[\onlinecite{planes_multical}], that a multicaloric effect is the result
  of applying/removing multiple fields, additional to the sum of
  applying/removing each field separately. The caloric response in one ferroic
  property due to the field conjugate to another ferroic property coupled to
  the first one is called a cross-caloric effect.}\BibitemShut {Stop}%
\bibitem [{cri()}]{critexpcomment}%
  \BibitemOpen
  \href@noop {} {}\bibinfo {note} {The precise critical exponents would of
  course be altered in a theory correctly considering
  fluctuations.}\BibitemShut {Stop}%
\end{thebibliography}
\bibliographystyle{apsrev4-1}

\end{document}

\title{Supplementary Material: Prediction of a Giant Magnetoelectric Cross-Caloric Effect Around a Tetracritical Point in Multiferroic SrMnO$_3$}
 
\author{Alexander Edstr\"om}
\author{Claude Ederer}
\affiliation{Materials Theory, ETH Z\"urich, Wolfgang-Pauli-Strasse 27, 8093 Z\"urich, Switzerland}

\begin{abstract}
This supplementary material contains a detailed description of the procedure for determining all parameters entering the Landau theory type free energy used in the paper. Additionally, it contains a table with the parameters, a summary of some thermodynamic relations used, as well as the phonon specific heat as function of temperature evaluated from frozen phonon calculations. 
\end{abstract}
%\date{\today}

\maketitle

\section{First Principles Computational Methods}

All density functional theory (DFT) calculations are performed as in Ref.~[\onlinecite{PhysRevMaterials.2.104409}], i.e. with VASP~\cite{KRESSE199615,PhysRevB.49.14251,PhysRevB.47.558} and projector augmented wave (PAW) pseudopotentials~\cite{PhysRevB.50.17953,PhysRevB.59.1758}. The exchange-correlation functional is described with the PBEsol version of the generalized gradient approximation~\cite{PhysRevLett.100.136406} (GGA). A Coulomb repulsion~\cite{PhysRevB.57.1505} of $U_\mathrm{eff}=3~\mathrm{eV}$ is included on the Mn $d$-electrons. The phonon specific heat was obtained from frozen phonon calculations performed with the Phonopy software~\cite{TOGO20151}. 

\section{Landau Theory for {S\MakeLowercase{r}M\MakeLowercase{n}O$_3$}}

We consider a Landau free energy of the form 
\begin{equation}\label{eq.LandauF}
	\mathcal{F}_q =   \frac{1}{2} a_P(T,\eta) P^2 + \frac{b_P}{4} P^4 + \frac{1}{2} a_q(T,\eta) M_q^2 + \frac{b_q}{4} M_q^4  + \frac{\lambda_q(\eta)}{2} M_q^2P^2  - EP ,
\end{equation}
for each magnetic order $q$, with strain and temperature dependence entering as 
\begin{equation}
a_q = \alpha_q (T - \tzq) + c_q \eta 
\end{equation}
\begin{equation}
a_P = \alpha_P (T - \tzp) + c_P \eta,
\end{equation}
with $T$ denoting temperature, $T_{0}^i$\footnote{We use the notation that $T_{c}^i$ is the strain and field dependent critical temperature of order parameter $i$, while $T_{0}^i$ is the value at zero strain and field.} the critical temperature at zero strain and applied field for the order parameter $i$, $P$ is the electric polarization and $M_q = \frac{1}{N} \sum_j^N \ee^{\img \mathbf{q} \cdot \mathbf{R}_j} \langle S_j \rangle$ is the magnetic order parameter for the order labelled $q$, with $\langle S_j \rangle$ denoting the thermodynamic average of the $j$th unitless, normalized spin $S_j$ out of $N$, and $\eta$ denotes (biaxial tensile) strain. $a_i$ and $b_i$ are the quadratic and quartic coefficients for the order parameter $i$, $c_i$ its coupling to strain and $\lambda_q$ the coupling between $P$ and the magnetic order parameter labeled by $q$. We consider magnetic order parameters corresponding to ferromagnetism (F) as well as G, C, and A-type antiferromagnetism. 

In the following we describe in detail how all the parameters entering the Landau free energy in Eq.~\eqref{eq.LandauF} are determined.
We first consider the uncoupled case ($\lambda_q=0$), in which ferroelectricity and magnetism can be considered separately. This is discussed in Secs.~\ref{sec.FE}-\ref{sec.mag}, respectively. The calculation of the coupling constant $\lambda_q$ is then discussed in Sec.~\ref{sec.coup}, after which the zero-field susceptibilities and thermodynamics of the given Landau theory is discussed in Sec.~\ref{sec.susc_therm}.

\subsection{Ferroelectricity}\label{sec.FE}

The ferroelectricity, without coupling to magnetism and with no applied field, is described by a free energy 
\begin{equation}\label{P_F}
	F_P =  \frac{1}{2}\left[ \alpha_P(T-\tzp) + c_P \eta \right] P^2 + \frac{b_P}{4} P^4, 
\end{equation}

with four parameters, $\alpha_P$, $\tzp$, $c_P$ and $b_P$, to be determined. Some of these could be determined by fitting total energy DFT calculations for fixed atomic displacements, corresponding to the ferroelectric (FE) soft mode displacement amplitude $u$ (see Ref.~\onlinecite{PhysRevMaterials.2.104409}), and fitting a curve $E(u)=E_0 + a_2 u^2 + a_4 u^4$, where the polarization $P$ is proportional to $u$ via the Born effective charges $Z^*$. Such fittings are essentially presented in Ref.~[\onlinecite{PhysRevMaterials.2.104409}]. Instead of using such a fitting procedure here, we notice that we need any four suitable pieces of information to fit four unknown parameters, by solving a linear system of equations. The following four are chosen: two parameters, namely $\tzp$ and the quotient $-c_P / \alpha_P$, are obtained from the fitting of $\tcp(\eta) = \tzp - \frac{c_P}{\alpha_P} \eta$ to the results of the {\emph{first principles}}-based effective Hamiltonian calculations of Ref.~[\onlinecite{PhysRevMaterials.2.104409}], where the FE critical temperature is found to be essentially linear in strain. The parameters $c_P$ and $\alpha_P$ are separately determined by setting $c_P=B_\mathrm{1xx} + B_\mathrm{1yy}(1-2B_{12}/B_{11})$, where the parameters $B_\mathrm{1xx}$, $B_\mathrm{1yy}$, $B_{12}$ and $B_{11}$ are taken from Ref.~[\onlinecite{PhysRevMaterials.2.104409}]. Finally $b_P$ is set to reproduce the DFT saturation polarization at 5\% strain at zero temperature using $P^2(T=0,\eta=0.05)=(-\alpha_P\tzp + 0.05c_P)/b_P$. 

\begin{figure}[hbt!]
	\centering
	\includegraphics[width=0.45\textwidth]{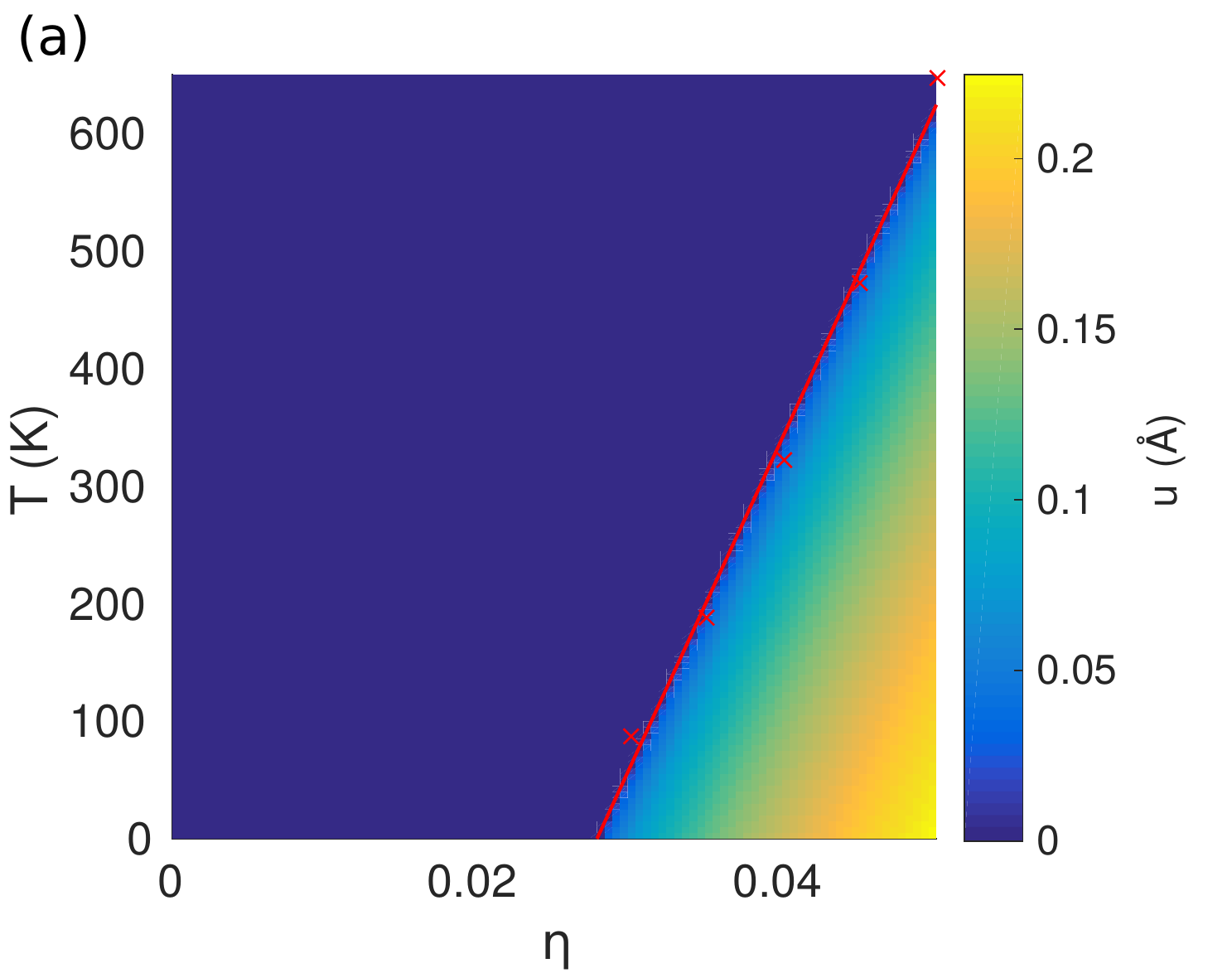}
	\includegraphics[width=0.45\textwidth]{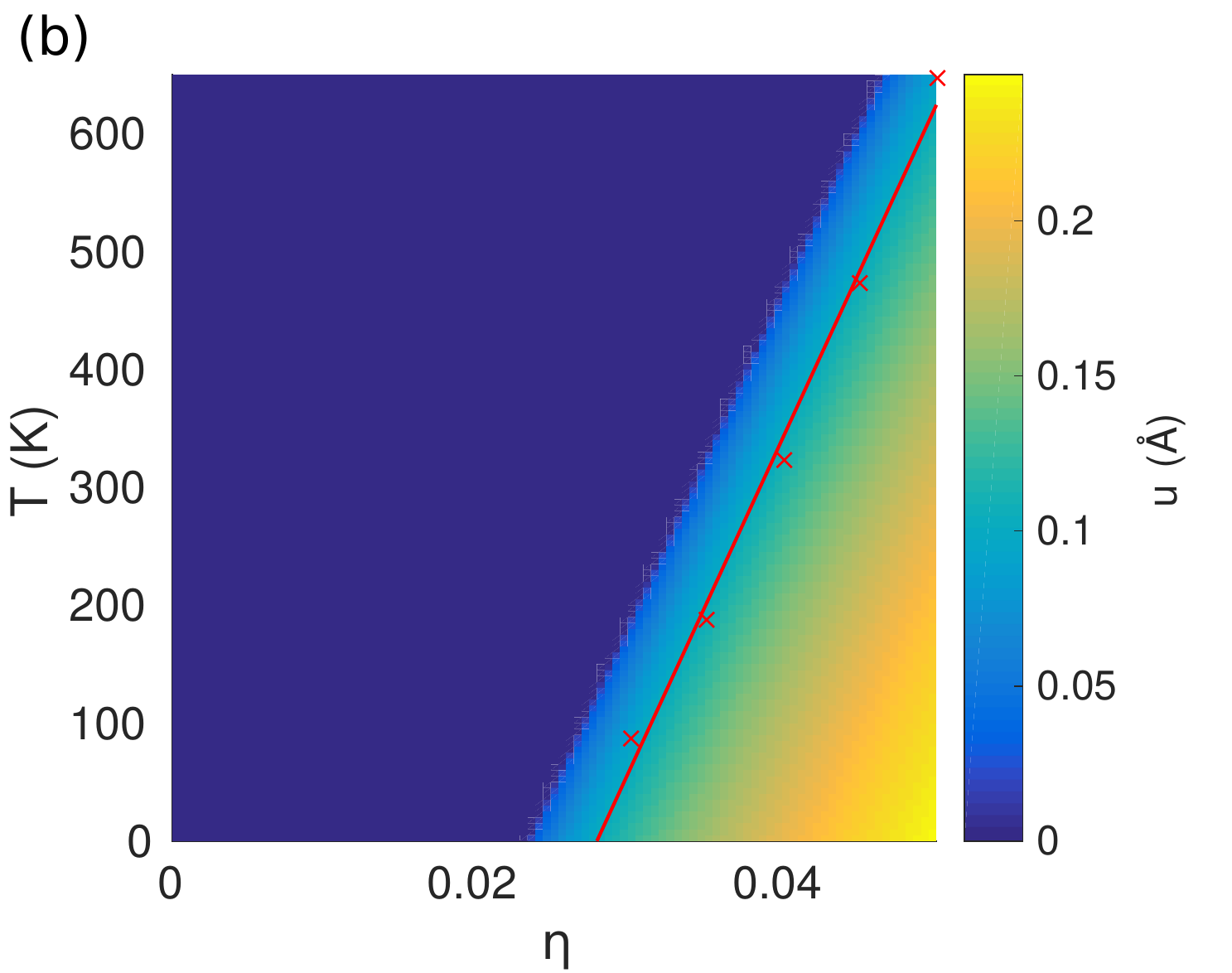}
	\caption{FE soft mode displacement $u \propto P$ as function of strain and temperature from the Landau theory with parameters obtained before (a) and after (b) correcting for the ME coupling, as discussed in the text. The red crosses and lines show the critical temperatures from the effective Hamiltonian approach of Ref.~[\onlinecite{PhysRevMaterials.2.104409}] and the corresponding linear fit.}
	\label{fig.P}
\end{figure}

With the resulting parameter values, we obtain the FE phase diagram shown in Fig.~\ref{fig.P}(a), which precisely reproduces the behavior of the effective Hamiltonian of Ref.~\onlinecite{PhysRevMaterials.2.104409} in terms of critical temperatures as function of strain and saturation polarization at $\eta=5\%$. However, all parameters of the effective Hamiltonian were obtained from DFT calculations for the G-type antiferromagnetic (AFM) state, and thus the effect of the coupling between G-type AFM order and the ferroelectricity needs to be subtracted. According to Eq.~\eqref{eq.LandauF} this coupling will contribute a term $\frac{\lambda_G}{2}M_\text{G}^2P^2$ to $F_P$ compared to Eq.~\eqref{P_F}, which will shift the zero strain critical temperature by $\lambda_G M_\text{G}^2/\alpha_P$ and modify $b_P$ by $-\lambda_G M_\text{G}^2/P_s^2(T=0,\eta=0.05)$, with $M_\text{G}^2=1$ being the fully saturated G-type AFM order parameter at $T=0$. The ferroelectric phase diagram, with corrected parameters $\tcp$ and $b_P$ (using the value obtained for $\lambda_G$ as described in Sec.~\ref{sec.coup}) is shown in Fig.~\ref{fig.P}(b). The main effect of correcting for the magnetoelectric (ME) coupling is a shift in the critical temperature as function of strain, and corresponding changes in the critical strain, as well as saturation polarization values. 

\subsection{Magnetic order}\label{sec.mag}

The magnetic order, excluding the coupling to the polarization, is described by the free energy 
\begin{equation}\label{Fmag}
F_q =  \frac{1}{2}\left[ \alpha_q(T-\tzq) + c_q \eta\right] M_q^2 + \frac{b_q}{4} M_q^4 . 
\end{equation}
The equilibrium magnetic phase at a given strain and temperature is obtained by minimizing each $F_q$ with respect to $M_q$ and the free energy is then given by $F_M = \min_{q} \left( \min_{M_q} F_q \right)$, where the $q$ corresponding to the minimum free energy describes the equilibrium magnetic phase at that point of the phase diagram. Each magnetic order parameter is characterized by the parameters $\alpha_q$, $\tzq$, $c_q$ and $b_q$. These will be determined via a mapping of the magnetic energies to a Heisenberg Hamiltonian
\begin{equation}\label{eq.Heis}
E = -\frac{1}{2} \sum_{i,j} J_{ij} \mathbf{S}_i \cdot \mathbf{S}_j .
\end{equation}

\begin{figure}[hbt!]
	\centering
	\includegraphics[width=0.5\textwidth]{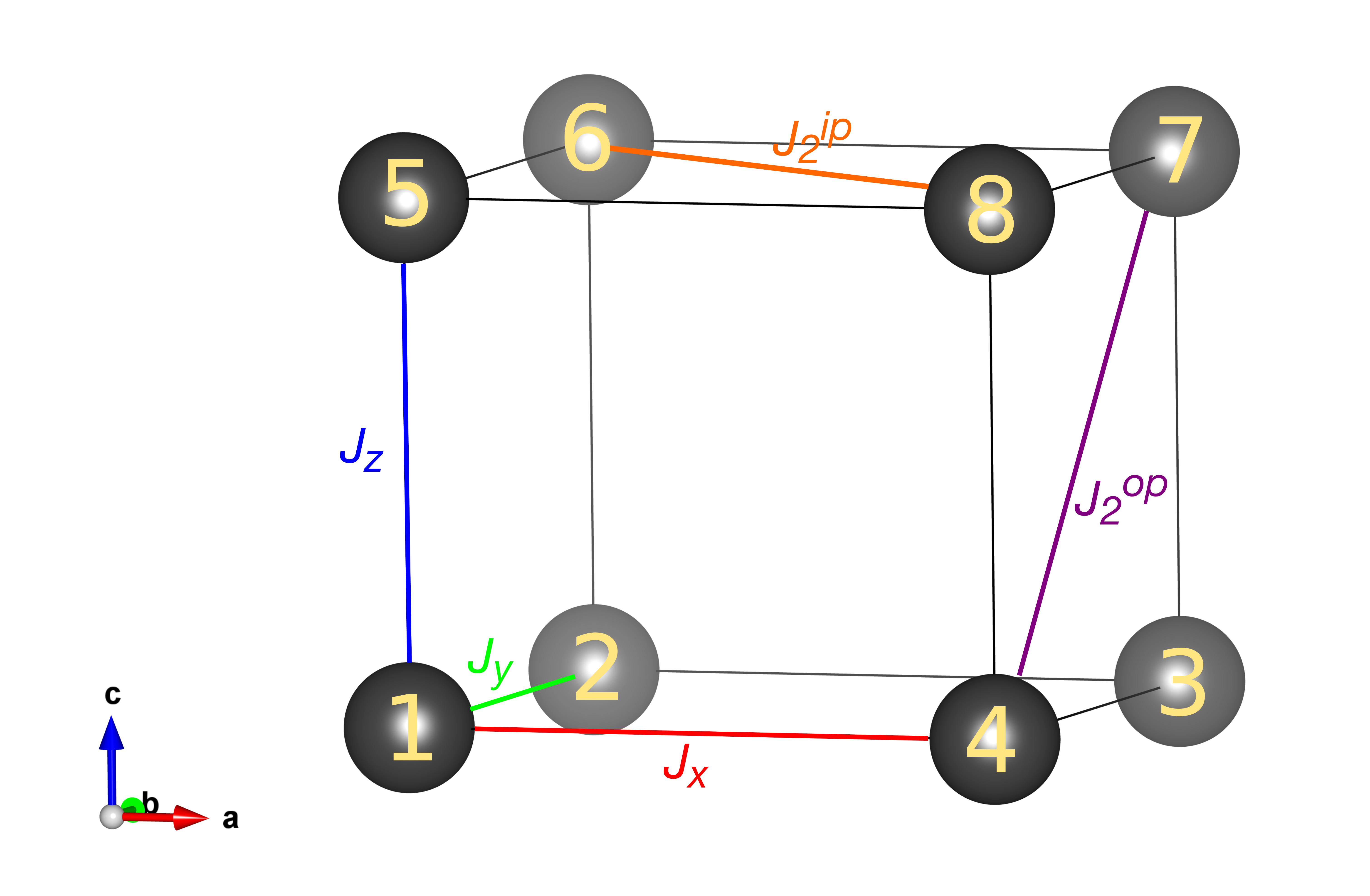}
	\caption{$2\times2\times 2$ cubic (or tetragonal in the strained case) perovskite supercell, considering only the B-site (Mn) atoms, which form a simple cubic (tetragonal) lattice. }
	\label{fig.supercell}
\end{figure}

Fig.~\ref{fig.supercell} illustrates the three nearest neighbor exchange interactions $J_x$, $J_y$ and $J_z$ for the coupling between spins on Mn atoms located relative to each other in the $x$, $y$ and $z$-directions respectively. In the cubic structure these are all equivalent $J_x=J_y=J_z=J_1$. With biaxial tensile strain these are split into two inequivalent interactions, in plane (ip) $J_x=J_y=J_1^\mathrm{ip}$ and out-of-plane (op) $J_z=J_1^\mathrm{op}$. Considering also polar displacements can make all three interactions inequivalent. Also the second nearest neighbor interactions $J_2$ are all equivalent in the cubic phase, while biaxial tensile strain results in two inequivalent couplings, in-plane or out-of-plane, as shown in Fig.~\ref{fig.supercell}. The second nearest neighbor interactions are, however, only considered fixed at the values for the cubic structure, since they are small and do not change sign with strain or polar distortions. They must, nevertheless, be included at least in the cubic structure in order to stabilize C-type AFM order over A-type in some strain range, as predicted by the DFT calculations to which these parameters are fitted. 

The magnetic order parameters are defined as 
\begin{equation}
\label{eq.Mq}
M_\mathbf{q} = \frac{1}{N} \sum_j^N	\ee^{\img \mathbf{q} \cdot \mathbf{R}_j} \langle S_j \rangle ,
\end{equation}
in terms of the $N=8$ spins, corresponding to the Mn sites inside a $2\times 2\times 2$ supercell of the SrMnO$_3$ structure, as shown in Fig.~\ref{fig.supercell}.
This cell is compatible with all relevant magnetic order parameters, including ferromagnetic [$\mathbf{q}=(0,0,0)$], G-type AFM [$\mathbf{q}=(1,1,1)$], C-type AFM [$\mathbf{q}=(1,1,0)$, $(1,0,1)$, or $(0,1,1)$], and A-type AFM [$\mathbf{q}=(0,0,1)$, $(0,1,0)$, or $(1,0,0)$] order.
Here, all wave-vectors are given in units of $\pi$ divided by the corresponding real space lattice constant.
Within cubic symmetry, all three $\mathbf{q}$-vectors corresponding to C-type or A-type AFM order, respectively, are equivalent and thus energetically degenerate. When we consider biaxial tensile strain, resulting in tetragonal symmetry, however, $(1,1,0)$ is different from $(1,0,1)$ and $(0,1,1)$. Similarly, for A-type, $(0,0,1)$ becomes inequivalent to $(1,0,0)$ and $(0,1,0)$. Introducing also a polarization, results in additional symmetry breaking. Hence, we will initially consider all eight $\mathbf{q}$-vectors listed above, that is one each for ferromagnetic and G-type AFM orders, and three each for A and C-type antiferromagnetism. 

Specifically, for these 8 order parameters, Eq.~\eqref{eq.Mq} becomes:
\begin{align}\label{eq.ordpar1}
M_{000} & = \frac{1}{8}\left( \langle S_1\rangle  + \langle S_2\rangle  + \langle S_3\rangle  + \langle S_4\rangle  + \langle S_5\rangle  + \langle S_6\rangle  + \langle S_7\rangle  + \langle S_8\rangle  \right) \\ 
M_{111} & = \frac{1}{8}\left( \langle S_1\rangle  - \langle S_2\rangle  + \langle S_3\rangle  - \langle S_4\rangle  - \langle S_5\rangle  + \langle S_6\rangle  - \langle S_7\rangle  + \langle S_8\rangle  \right)\\
M_{110} & = \frac{1}{8}\left( \langle S_1\rangle  - \langle S_2\rangle  + \langle S_3\rangle  - \langle S_4\rangle  + \langle S_5\rangle  - \langle S_6\rangle  + \langle S_7\rangle  - \langle S_8\rangle  \right)\\
M_{101} & = \frac{1}{8}\left( \langle S_1\rangle  + \langle S_2\rangle  - \langle S_3\rangle  - \langle S_4\rangle  - \langle S_5\rangle  - \langle S_6\rangle  + \langle S_7\rangle  + \langle S_8\rangle  \right)\\
M_{011} & = \frac{1}{8}\left( \langle S_1\rangle  - \langle S_2\rangle  - \langle S_3\rangle  + \langle S_4\rangle  - \langle S_5\rangle  + \langle S_6\rangle  + \langle S_7\rangle  - \langle S_8\rangle  \right)\\
M_{001} & = \frac{1}{8}\left( \langle S_1\rangle  + \langle S_2\rangle  + \langle S_3\rangle  + \langle S_4\rangle  - \langle S_5\rangle  - \langle S_6\rangle  - \langle S_7\rangle  - \langle S_8\rangle  \right) \\
M_{010} & = \frac{1}{8}\left( \langle S_1\rangle  - \langle S_2\rangle  - \langle S_3\rangle  + \langle S_4\rangle  + \langle S_5\rangle  - \langle S_6\rangle  - \langle S_7\rangle  + \langle S_8\rangle  \right) \\
M_{100} & = \frac{1}{8}\left( \langle S_1\rangle  + \langle S_2\rangle  - \langle S_3\rangle  - \langle S_4\rangle  + \langle S_5\rangle  + \langle S_6\rangle  - \langle S_7\rangle  - \langle S_8\rangle  \right)
\label{eq.ordpar8}
\end{align}
or $\mathbf{M} = V \mathbf{S}$ with 
\begin{equation} \label{eq.Pmat}
V = \frac{1}{8} \begin{pmatrix}
1 & ~~1 & ~~1 & ~~1 & ~~1 & ~~1 & ~~1 & ~~1 \\
1 & -1 & ~~1 & -1 & -1 & ~~1 & -1 & ~~1 \\
1 & -1 & ~~1 & -1 & ~~1 & -1 & ~~1 & -1 \\
1 & ~~1 & -1 & -1 & -1 & -1 & ~~1 & ~~1 \\
1 & -1 & -1 & ~~1 & -1 & ~~1 & ~~1 & -1 \\
1 & ~~1 & ~~1 & ~~1 & -1 & -1 & -1 & -1 \\
1 & -1 & -1 & ~~1 & ~~1 & -1 & -1 & ~~1 \\
1 & ~~1 & -1 & -1 & ~~1 & ~~1 & -1 & -1 \\ 
\end{pmatrix}
\quad \mathrm{and} \quad 
V^{-1}=8V^\mathrm{T}=\begin{pmatrix}
 1 & ~~1 & ~~1 & ~~1 & ~~1 & ~~1 & ~~1 & ~~1 \\
 1 & -1 & -1 & ~~1 & -1 & ~~1 & -1 & ~~1 \\
 1 & ~~1 & ~~1 & -1 & -1 & ~~1 & -1 & -1 \\
 1 & -1 & -1 & -1 & ~~1 & ~~1 & ~~1 & -1 \\
 1 & -1 & ~~1 & -1 & -1 & -1 & ~~1 & ~~1 \\
 1 & ~~1 & -1 & -1 & ~~1 & -1 & -1 & ~~1 \\
 1 & -1 & ~~1 & ~~1 & ~~1 & -1 & -1 & -1 \\
 1 & ~~1 & -1 & ~~1 & -1 & -1 & ~~1 & -1 \\
\end{pmatrix} .
\end{equation}
Here $\langle S_i\rangle $ denotes the thermal average of spin $S_i$ on site $i$, projected on the spin quantization axis. Using $V^{-1}$, Eq.~\eqref{eq.ordpar1}-\eqref{eq.ordpar8} can be inverted to
\begin{align}\label{eq.spinordpar1}
\langle S_1\rangle   & = M_{000} + M_{111} + M_{110} + M_{101} + M_{011} + M_{001} + M_{010}+ M_{001} \\
\langle S_2\rangle   & = M_{000} - M_{111} - M_{110} + M_{101} - M_{011} + M_{001} - M_{010}+ M_{001} \\
\langle S_3\rangle   & = M_{000} + M_{111} + M_{110} - M_{101} - M_{011} + M_{001} - M_{010}- M_{001} \\
\langle S_4\rangle   & = M_{000} - M_{111} - M_{110} - M_{101} + M_{011} + M_{001} + M_{010}- M_{001} \\
\langle S_5\rangle   & = M_{000} - M_{111} + M_{110} - M_{101} - M_{011} - M_{001} + M_{010}+ M_{001} \\
\langle S_6\rangle   & = M_{000} + M_{111} - M_{110} - M_{101} + M_{011} - M_{001} - M_{010}+ M_{001} \\
\langle S_7\rangle   & = M_{000} - M_{111} + M_{110} + M_{101} + M_{011} - M_{001} - M_{010}- M_{001} \\
\langle S_8\rangle   & = M_{000} + M_{111} - M_{110} + M_{101} - M_{011} - M_{001} + M_{010}- M_{001} .
\label{eq.spinordpar8}
\end{align}

In the mean field approximation, the spins in Eq.~\eqref{eq.Heis} can be substituted with their thermal averages. 
\begin{equation}\label{eq.Heis_mf}
E =  -\frac{1}{2} \sum_{i,j} J_{ij} \mathbf{S}_i \cdot \mathbf{S}_j \approx -\frac{1}{2} \sum_{i,j} J_{ij} \langle S_i\rangle  \cdot \langle S_j\rangle 
\end{equation}
Considering a cubic structure with first and second nearest neighbor interactions, substituting Eq.~\eqref{eq.spinordpar1}-Eq.~\eqref{eq.spinordpar8} into Eq.~\eqref{eq.Heis_mf} yields  
\begin{align}\label{eq.Heis_E_of_ordpar}
\frac{E}{8} = &  - \frac{1}{4} J_x (\langle S_1\rangle  \langle S_4\rangle  + \langle S_2\rangle  \langle S_3\rangle  +  \langle S_5\rangle  \langle S_8\rangle  + \langle S_6\rangle  \langle S_7\rangle ) -  \nonumber \\
	  &  - \frac{1}{4} J_y  (\langle S_1\rangle  \langle S_2\rangle  + \langle S_3\rangle  \langle S_4\rangle  +  \langle S_5\rangle  \langle S_6\rangle  + \langle S_7\rangle  \langle S_8\rangle ) - \nonumber \\
	 &   - \frac{1}{4} J_z  (\langle S_1\rangle  \langle S_5\rangle  + \langle S_2\rangle  \langle S_6\rangle  +  \langle S_3\rangle  \langle S_7\rangle  + \langle S_4\rangle  \langle S_8\rangle ) = \nonumber \\
	 & - \frac{1}{2} J_2 ( \langle S_1\rangle \langle S_3\rangle  + \langle S_2\rangle \langle S_4\rangle  + \langle S_5\rangle \langle S_7\rangle  + \langle S_6\rangle \langle S_8\rangle   + \nonumber \\ 
	 & + \langle S_1\rangle  \langle S_6\rangle  + \langle S_2\rangle  \langle S_5\rangle  + \langle S_4\rangle  \langle S_7\rangle  + \langle S_3\rangle  \langle S_8\rangle  + \nonumber \\ 
	 & + \langle S_1\rangle  \langle S_8\rangle  + \langle S_4\rangle  \langle S_5\rangle  + \langle S_2\rangle  \langle S_7\rangle  + \langle S_3\rangle  \langle S_6\rangle ) = \nonumber \\ 
	 = &  -(J_x + J_y + J_z + 6 J_2)M_{000}^2 + (J_x + J_y + J_z - 6J_2)M_{111}^2 + \nonumber \\ 
  	 & + (J_x + J_y - J_z + 2 J_2)M_{110}^2 + (J_x - J_y + J_z + 2 J_2)M_{101}^2 + (-J_x + J_y + J_z + 2 J_2)M_{011}^2 - \nonumber \\ 
  	 & -(J_x + J_y - J_z - 2 J_2)M_{001}^2 - (J_x - J_y + J_z - 2 J_2)M_{010}^2 - (-J_x + J_y + J_z - 2 J_2)M_{100}^2 ,
\end{align}
where the division by eight gives normalization per perovskite unit cell and $J_2$ is assumed to be the same in every direction. In the cubic structure, where also $J_1$ is the same in every direction, Eq.~\eqref{eq.Heis_E_of_ordpar} can be written 
\begin{equation}\label{eq.Heis_E_of_ordpar_cube}
\frac{E}{8} = \underbrace{-3(J_1 + 2J_2) }_{-\frac{1}{2}\alpha_FT_{0}^F}M_F^2 +  \underbrace{3(J_1 - 2J_2)}_{-\frac{1}{2}\alpha_GT_{0}^G}M_G^2 +   \underbrace{(J_1 + 2J_2)}_{-\frac{1}{2}\alpha_CT_{0}^C}M_C^2  \underbrace{-(J_1 - 2J_2)}_{-\frac{1}{2}\alpha_AT_{0}^A}M_A^2  ,
\end{equation}
Since the three different A or C-type AFM orders are now degenerate they were only included once each and labeled by the corresponding letter instead of $\mathbf{q}$-vector. By identifying the energy of the Heisenberg Hamiltonian with the free energy at zero temperature, the quadratic coefficients of the Landau free energy can be identified in terms of the Heisenberg exchange interactions in Eq.~\eqref{eq.Heis_E_of_ordpar_cube}. 

The critical temperatures at zero strain, for each of the magnetic order parameters, are obtained using multi-sublattice mean field theory~\cite{PhysRevB.70.024427,Anderson196399}. For a unit cell with $N$ magnetic atoms, an exchange matrix $\mathcal{J}$ is constructed as
\begin{equation}
[\mathcal{J}]_{AB} = \sum_i J_{A_0B_i},
\end{equation}
where $A$ and $B$ denote magnetic sublattices and $i$ is an index running over different atomic sites of type $B$. $\mathcal{J}$ is thus a symmetric $N \times N$ matrix. The critical temperature is
\begin{equation} \label{eq.mft_Tc}
T_\mathrm{c} = \frac{J_0}{3k_\mathrm{B}},
\end{equation} 
where $k_\mathrm{B}$ is Boltzmann's constant and $J_0$ is the largest eigenvalue of $\mathcal{J}$. The corresponding eigenvector describes the magnetic order. By looking at the other eigenvalues, information regarding the hypothetical critical temperatures of magnetic orders other than the most stable one can be obtained. In this manner, a critical temperature for each of the magnetic orders in the cubic structure is found to be 
\begin{equation}
\label{eq.Tcs}
T_{0}^F = -345.6~\mathrm{K} \quad , \quad T_{0}^G = 262.6~\mathrm{K} \quad  , \quad T_{0}^C = 115.2~\mathrm{K} \quad  , \quad T_{0}^A = -87.5~\mathrm{K}.
\end{equation}
The negative ordering temperatures indicate that the non-magnetic ($M_q=0$) solution is energetically favored relative to the corresponding magnetic order parameter being non-zero, in the cubic structure.

\begin{figure}[hbt!]
	\centering
	\includegraphics[width=0.49\textwidth]{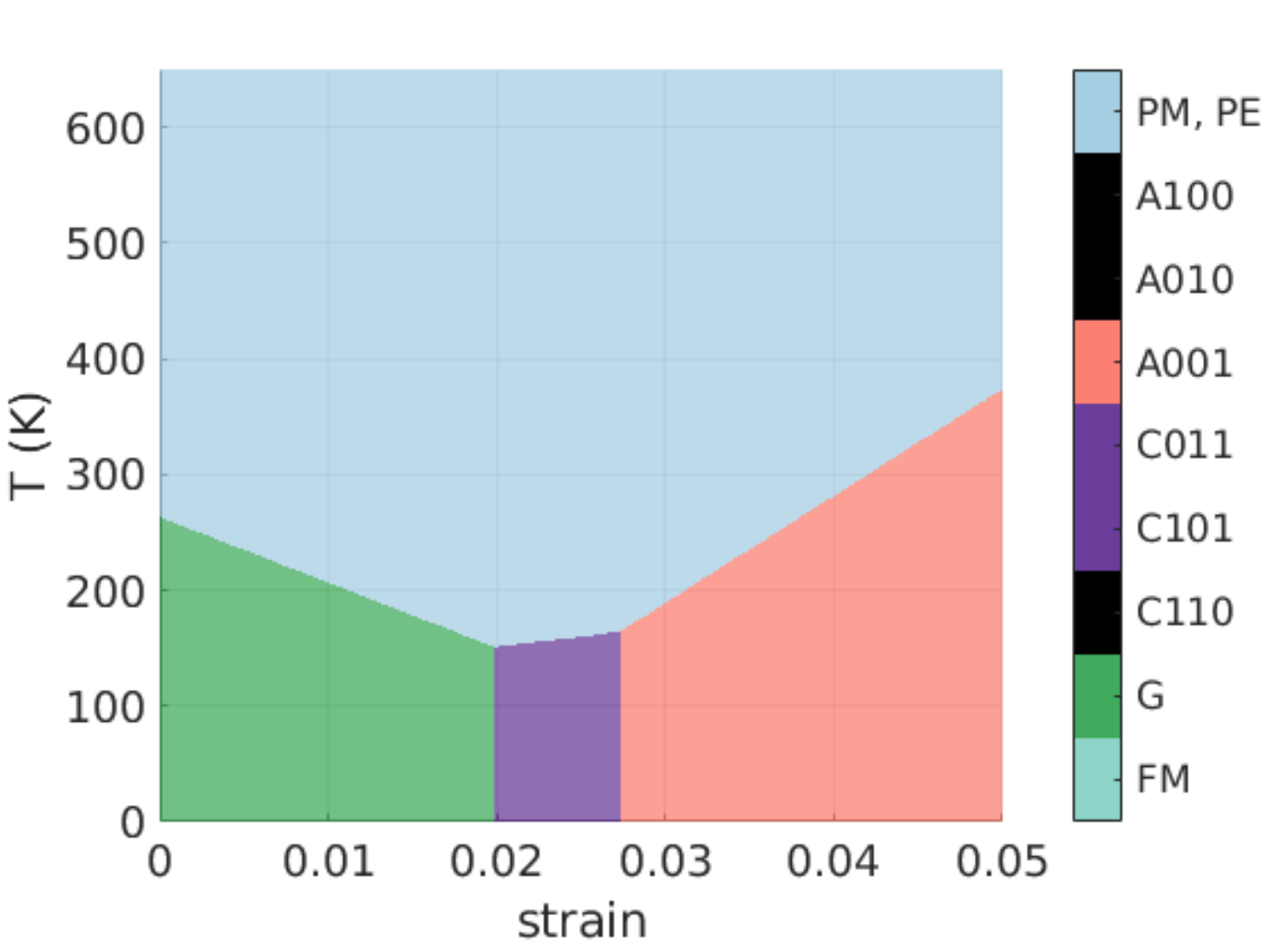} \includegraphics[width=0.49\textwidth]{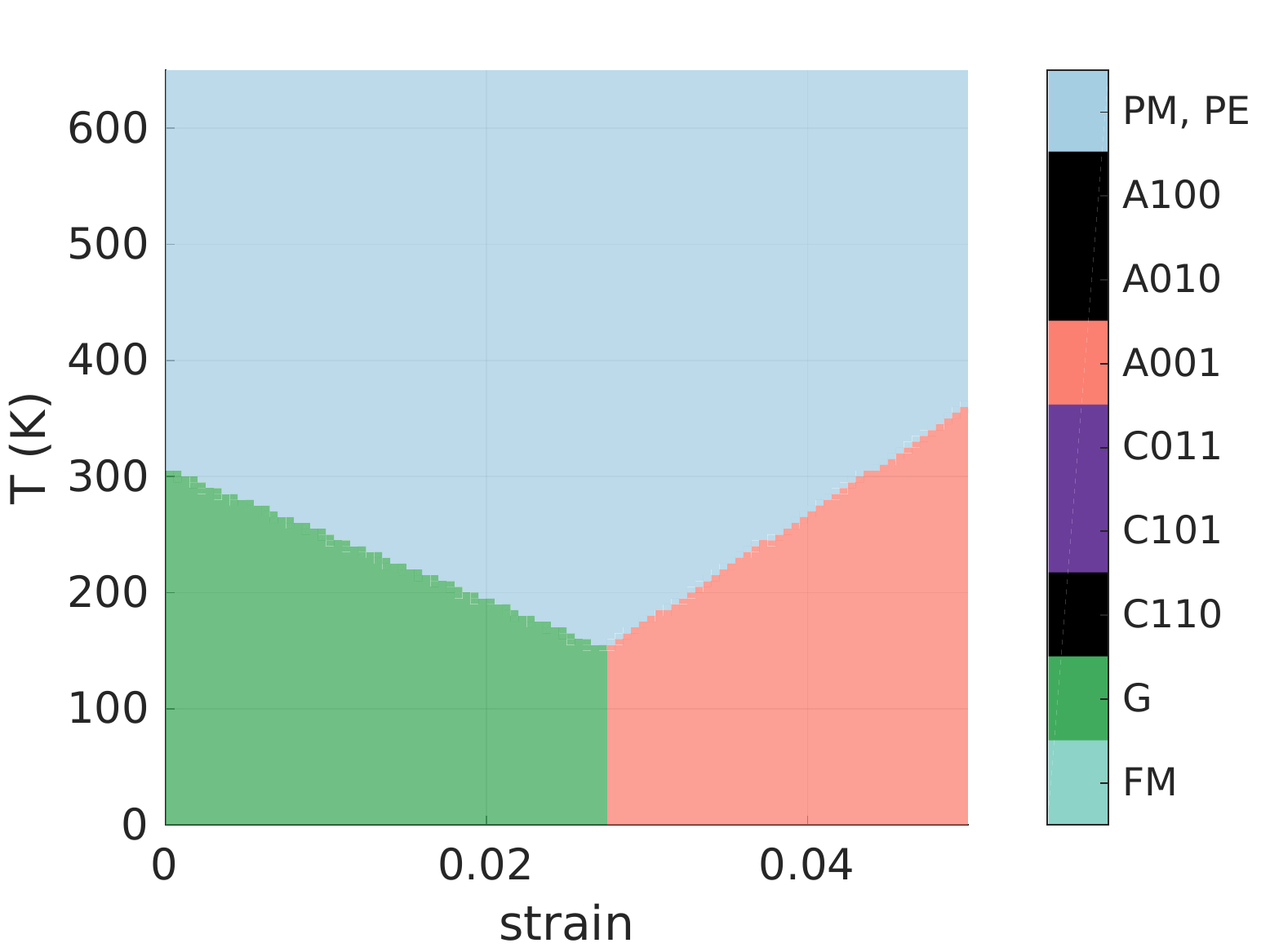}
	\caption{Magnetic phase diagram, excluding the coupling to the FE polarization (left). To the right is the result when neglecting $J_2$. In this case there is no region with C-type antiferromagnetism.}
	\label{fig.Tc_eta}
\end{figure}

The energy in Eq.~\eqref{eq.Heis_mf}, normalized per unit cell, can be written 
\begin{equation}
\frac{E}{8} = -\frac{1}{8} \frac{1}{2} \mathbf{S}^T \mathcal{J} \mathbf{S} = -\frac{1}{8} \frac{1}{2} \mathbf{M}^T (V^{-1})^T \mathcal{J} V^{-1} \mathbf{M} = - \frac{1}{2} \mathbf{M}^T \underbrace{V \mathcal{J} V^{-1}}_{D} \mathbf{M} = - \frac{1}{2} \mathbf{M}^T D \mathbf{M} ,
\end{equation}
where we used $(V^{-1})^T = 8V$, see Eq.~\eqref{eq.Pmat}, and $\mathbf{S}$ denotes an $N$-dimensional vector of $N$ spins, instead of a three-dimensional spin vector. Since no cross-coupling terms between different order parameters occur in Eq.~\eqref{eq.Heis_E_of_ordpar}, it is clear that $D$ must be diagonal and thus the change of variables defined by Eq.~\eqref{eq.ordpar1}-\eqref{eq.ordpar8} diagonalizes the spin Hamiltonian in Eq.~\eqref{eq.Heis_mf}. Thus, the matrix $D$ contains the eigenvalues of $\mathcal{J}$, i.e., $3k_\mathrm{B}$ times the corresponding ordering temperature (see Eq.~\eqref{eq.mft_Tc}), and these are also the coefficients of the squared order parameters in Eq.~\eqref{eq.Heis_E_of_ordpar}. From this it follows that $\alpha_q = 3k_\mathrm{B}$ for each $q$. One can also show that if two magnetic orders have equal $\alpha_q$ and there is a phase boundary as function of strain between them, the phase boundary will be vertical in the strain-temperature phase diagram (i.e., independent of temperature). Hence, all magnetic phase boundaries are vertical in the model used here, as is also seen in Fig.~\ref{fig.Tc_eta}. 

For a given order parameter, excluding any coupling to other order parameters, the value which minimizes the free energy is 
\begin{equation}
M_q^2 = -\frac{a_q}{b_q} .
\end{equation}
According to the definitions in Eq.~\eqref{eq.ordpar1}-\eqref{eq.ordpar8}, normalization of the spins implies that the zero temperature values of the order parameters will also be normalized to unity. Therefore, the quartic coefficients are set to the strain dependent values of 
\begin{equation}
b_q = -a_q (T=0). 
\end{equation}
The strain dependence of $b_q$ ensures that the zero temperature normalization of the spins is correct at all strains.

With all the necessary parameters determined, the magnetic phase diagram, excluding coupling to the FE polarization, can be obtained. Looking at the definitions of the magnetic order parameters in Eq.~\eqref{eq.ordpar1}-\eqref{eq.ordpar8}, one can deduce that the different magnetic order parameters exclude each other, i.e., one being unity implies that the others are zero. Furthermore, the adiabatic magnon spectra show minima only for zone center or zone boundary $\mathbf{q}$-vectors, at every strain considered~\cite{PhysRevMaterials.2.104409}, within the current model based on the Heisenberg Hamiltonian on a tetragonal lattice. Thus, the magnetic transitions are described by minimizing the total energy independently for each of the order parameters and taking the one that gives the lowest free energy as the only one non-zero for a given $(\eta,T)$. The resulting phase diagram is shown in Fig.~\ref{fig.Tc_eta}. The phase diagram obtained without the strain-independent $J_2$ is also shown to illustrate that $J_2$ is necessary to stabilize the C-type AFM region, which has been predicted by the DFT calculations which the parameters are fitted to.

\subsection{Coupling}\label{sec.coup}

\begin{figure}[hbt!]
	\centering
	\includegraphics[width=0.48\textwidth]{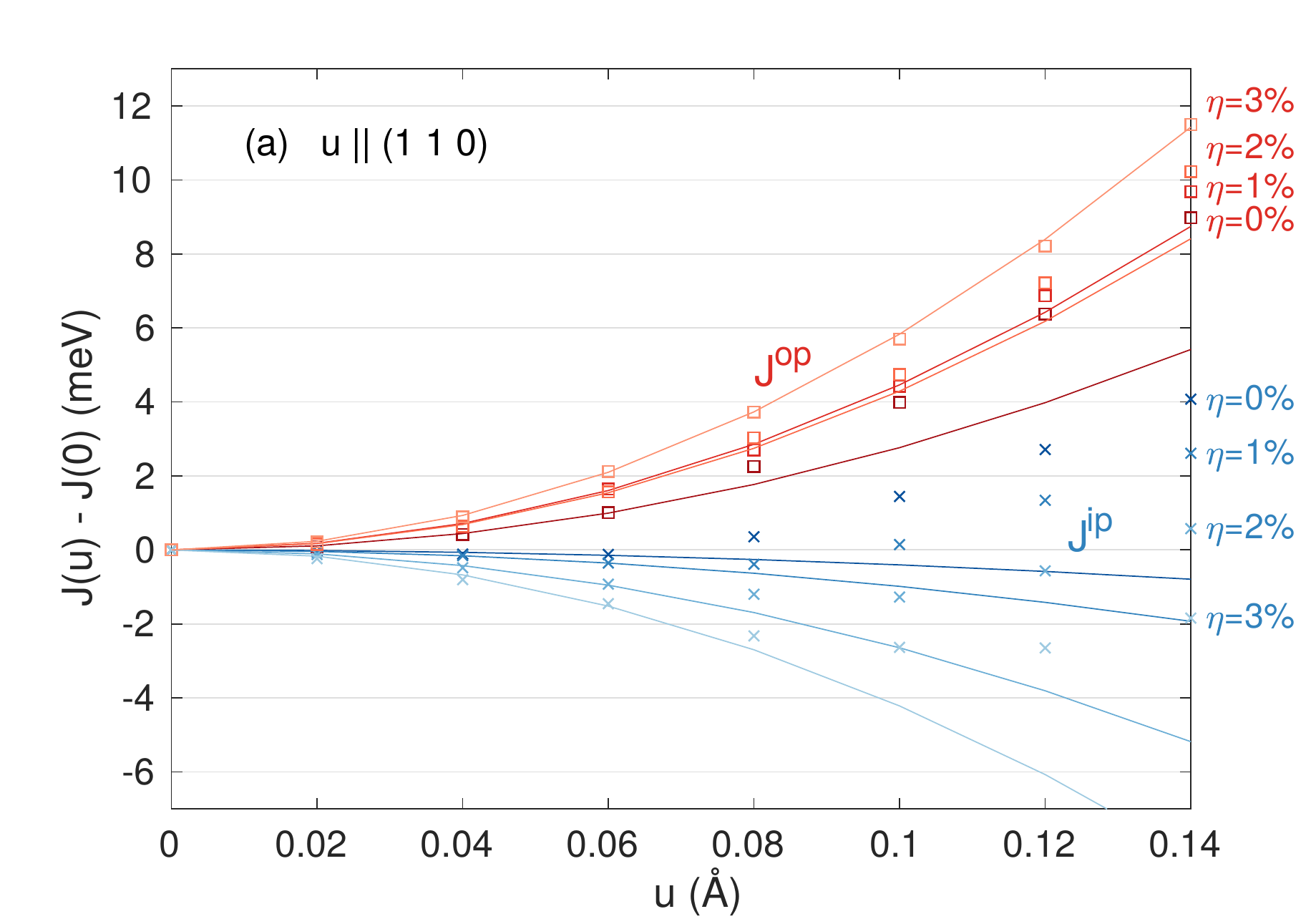}	
	\includegraphics[width=0.48\textwidth]{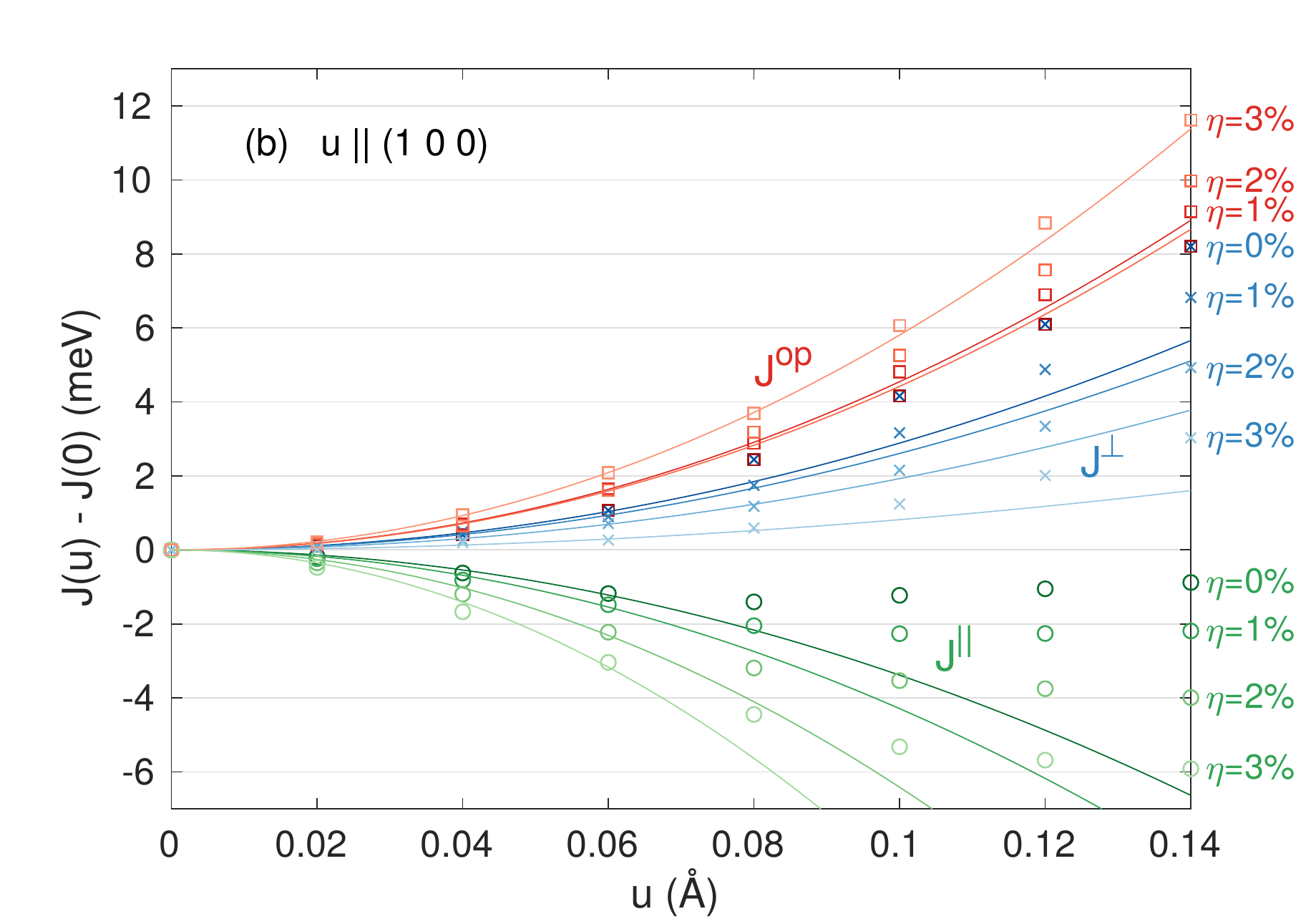}
	\caption{Change of the nearest neighbor exchange interactions (relative to their values for $u=0$) as function of FE soft mode amplitude $u$, for various strains $\eta$. The polarization is along the (110)-direction in (a) and along (100) in (b), while a 001-biaxial tensile strain is applied. Curves show quadratic fits of the data up to $u=0.06~\mathrm{\AA}$. }
	\label{fig.J_of_u}
\end{figure}

The last part needed before the complete multiferroic phase diagram can be established, are the biquadratic coupling coefficients $\lambda_q$. These are obtained by computing the magnetic exchange interactions as functions of a polarization $P$, by performing DFT calculations while freezing in atomic displacements according to a soft mode vector with amplitude $\mathbf{u} \propto \mathbf{P}$ . The computed exchange interactions as function of soft mode amplitude/polarization for various strains, is shown in Fig.~\ref{fig.J_of_u}, with (a) showing results for $\mathbf{u} \parallel (110)$ and (b) showing $\mathbf{u} \parallel (100)$. With biaxial tensile (001)-strain and polarization along the $(100)$-direction, all three nearest neighbor directions $J_x=J^\parallel$, $J_y=J^\perp$ and $J_z = J^\mathrm{op}$, are inequivalent, while with $\mathbf{u} \parallel (110)$ the two in-plane coupling parameters $J^\mathrm{ip}$ are equivalent. 

The biquadratic coupling between magnetism and ferroelectricity is obtained from a quadratic fit of the $J$'s as function of $u$. As can be seen in Fig.~\ref{fig.J_of_u}, this coupling is strongly strain dependent. Furthermore, it is not possible to produce a good quadratic fit for the whole range of relevant displacements/polarizations (which for large strain and low temperature can be larger than $u=0.2~\AA$). This is particularly clear for $J^\parallel$. A good fit over the whole range of temperature and strain considered, therefore, would require at least a higher order coupling term $\sim M^2 P^4$, and each coupling term needs to be strain dependent. However, this makes it necessary to consider also other higher order terms to guarantee stable solutions. This significantly increases the complexity and the number of fitting parameters of the model used. 

If one is interested in effects near the phase transitions, where the order parameters are small, it is sufficient to include only the biquadratic coupling term. From Fig.~\ref{fig.J_of_u} one can estimate that such a fit is good for polar displacements up to $\sim 0.06~\AA$, and it is to the data up to this point that the curves in Fig.~\ref{fig.J_of_u} have been fitted.

According to Refs.~\onlinecite{PhysRevMaterials.2.104409} and \onlinecite{PhysRevB.84.104440}, under tensile epitaxial strain $\mathbf{P} \parallel (110)$ for all magnetic order parameters except C-type, where instead $\mathbf{P} \parallel (100)$. Therefore, the coupling coefficient $\lambda_C$ is evaluated for  $\mathbf{P} \parallel (100)$, while all others are evaluated for $\mathbf{P} \parallel (110)$. 

Writing $J_x(P) = J_{x}(0) + j_x P^2$ and similarly for $y$, $z$, and inserting this into Eq.~\eqref{eq.Heis_E_of_ordpar}, leads to the terms proportional to $M_q^2 P^2$ needed to identify $\lambda_q$: 
\begin{align}
E_\mathrm{coupling} = & \frac{1}{2} [ \underbrace{-2(j_x + j_y + j_z)}_{\lambda_{000}}M_{000}^2 + \underbrace{2(j_x + j_y + j_z)}_{\lambda_{111}}M_{111}^2 + \nonumber \\ 
  	 & + \underbrace{2(j_x + j_y - j_z)}_{\lambda_{110}}M_{110}^2 + \underbrace{2(j_x - j_y + j_z)}_{\lambda_{101}}M_{101}^2 + \underbrace{2(-j_x + j_y + j_z)}_{\lambda_{011}}M_{011}^2 - \nonumber \\ 
  	 & \underbrace{-2(j_x + j_y - j_z)}_{\lambda_{001}}M_{001}^2 \underbrace{- 2(j_x - j_y + j_z)}_{\lambda_{010}}M_{010}^2 \underbrace{- 2(-j_x + j_y + j_z)}_{\lambda_{100}}M_{100}^2 ] P^2.
\end{align}

\begin{figure}[hbt!]
	\centering
	\includegraphics[width=0.45\textwidth]{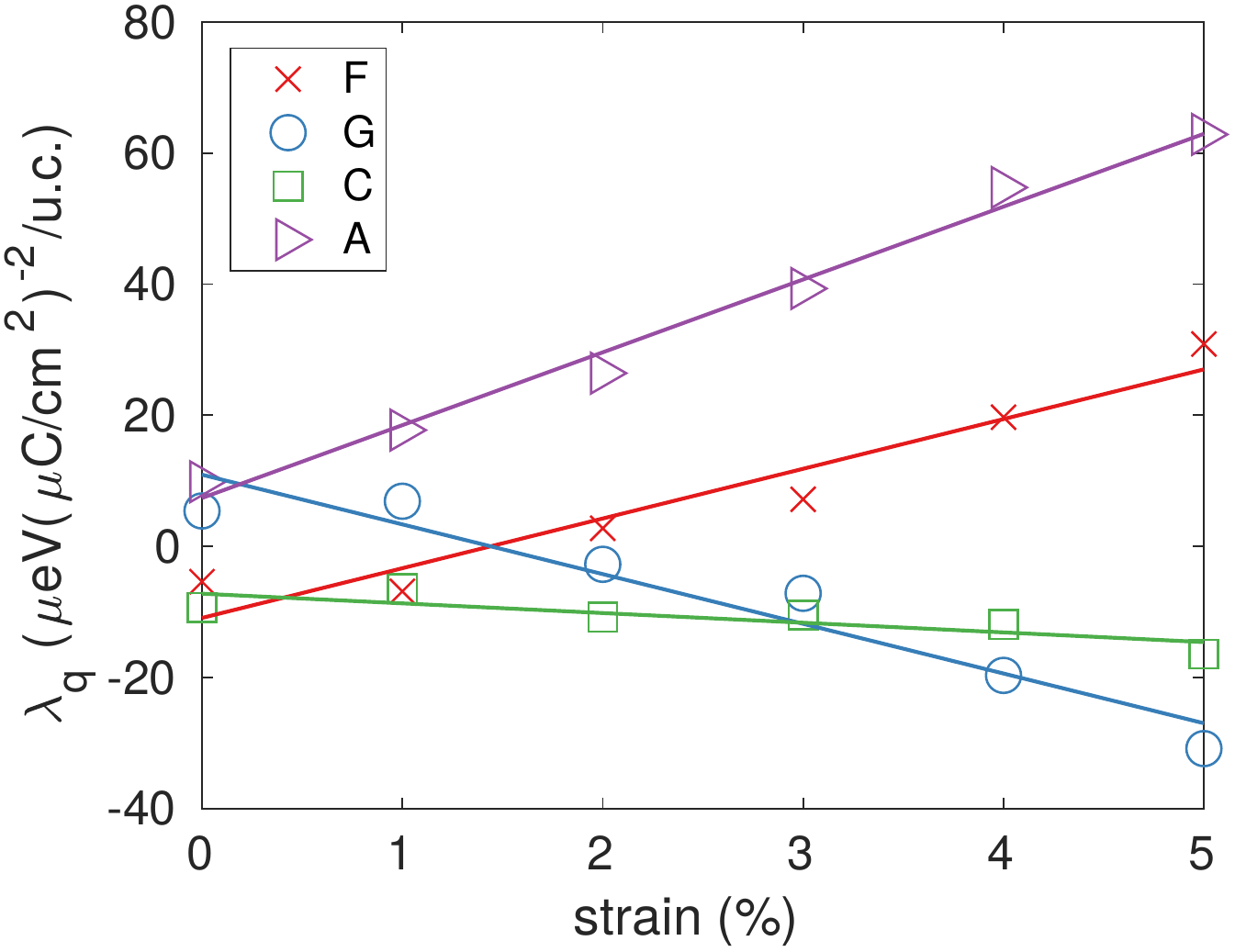}
	\caption{ Biquadratic magnetoelectric coupling parameters, $\lambda_q$, for different magnetic order parameters as functions of strain. The lines show linear fits to the data. }
	\label{fig.lambda_of_eta}
\end{figure}

By computing and fitting $J_x(\eta, P) = J_{x}(\eta,0) + j_x(\eta) P^2$, and similarly for $y$ and $z$, for various $\eta$, strain-dependent coupling parameters $\lambda_q(\eta)$ are obtained, as shown in Fig.~\ref{fig.lambda_of_eta}. Only the coupling parameters for the magnetic order parameters observed in the phase diagram are shown, i.e., A-type AFM order with $\mathbf{q}=(0,0,1)$ and C-type AFM order with $\mathbf{q}=(1,0,1)$. The coupling is to a polarization along the (110)-direction for all the magnetic orders except C, for which the polarization is along (100). The calculated $\lambda_q(\eta)$ are reasonably well described by a linear strain dependence, whereby it could be appropriate to describe the whole strain dependent phase diagram using linearly fitted coupling parameters. In this work, however, where focus is on the region $2\% < \eta < 3\%$,  and only the C-type AFM coupling is relevant, a linear interpolation between only these two data points is used. 

While the relevant C-type coupling parameter is negative for all strains and increases slightly in magnitude with increasing strain, the G-type coupling changes sign from positive to negative. At low strains, the G-type coupling is positive, which is consistent with previous suggestions that in cubic G-type AFM Sr$_{1-x}$Ba$_{x}$MnO$_3$ electric polarization and magnetic order disfavour each other~\cite{PhysRevLett.107.137601,PhysRevLett.109.107601}. 

\subsection{Summary of parameters}\label{sec.partable} 

Table~\ref{tab.param} contains the numerical values of all parameters entering the free energy density in Eq.~\eqref{eq.LandauF}, used in this work, except the strain dependent coupling parameters contained in Fig.~\ref{fig.lambda_of_eta}. 

\begin{table}[]
\caption{Values used for all parameters entering the expression for the free energy (u.c. = ``simple perovskite unit cell'').
} \begin{tabular}{l|ccccccccc|cc}
\hline \hline 
         $i$ & $M_{000}$ & $M_{111}$ & $M_{110}$ & $M_{101}$ & $M_{011}$ & $M_{001}$ & $M_{010}$ & $M_{100}$ & & $P$ & \\ \hline
$\alpha_i$ & $3$  &  $3$ &  $3$ & $3$ & $3$ & $3$ & $3$ & $3$ & $k_\mathrm{B}$ & $7.8\cdot 10^{-3}$ & \si{eV\angstrom^{-2}\kelvin^{-1}/{u.c.}} \\
$T_0^i$  & -345.6 & 262.6 & 115.2 & 115.2 & 115.2 & -87.5  & -87.5 & -87.5 & \si{\kelvin} & -575.7 & \si{\kelvin} \\
$c_i$ & -1.46  & 1.46  & 2.39  & -0.46 & -0.46  & -2.39  & 0.46  & 0.46 & \si{eV/{u.c.}} & -218.8 & \si{eV\angstrom^{-2}/{u.c.}} \\
$b_i$  & 8.94 & 6.79 & 2.98 & 2.98 & 2.98 & 2.26 & 2.26 & 2.26 & \si{10^{-2}}{eV/{u.c.}}& 96.4 & \si{eV\angstrom^{-4}/{u.c.}} \\ \hline \hline
\end{tabular}
\label{tab.param}
\end{table}

\subsection{Susceptibilities and thermodynamics}\label{sec.susc_therm}

We consider again a free energy such as that in Eq.~\eqref{eq.LandauF} but limit ourselves to one magnetic order parameter $M$. This is sufficient since the magnetic phase does not change within the region of the phase diagram of interest within this work. The equilibrium $M$ and $P$ fulfill
\begin{equation}\label{eq.Mcond}
\frac{\partial F}{\partial M} = M \left[ a_M + b_M M^2 + \lambda P^2 \right] = 0 \quad \rightarrow \quad M=0 \quad \mathrm{or} \quad M^2 = -\frac{1}{b_M} (a_M + \lambda P^2)
\end{equation}
and
\begin{equation}\label{eq.Pcond}
\frac{\partial F}{\partial P} = P \left[ a_P + b_P P^2 + \lambda M^2 \right] - E = 0.
\end{equation}
For $E=0$ the above conditions can easily be solved to produce the four different solutions in the top of Table~\ref{tab.E0Landau}. Taking derivatives of Eqs.~\eqref{eq.Mcond}-\eqref{eq.Pcond} with respect to $E$ and setting $E=0$, yields the zero field electric susceptibility
\begin{equation}
\chi_E =  \frac{\partial P}{\partial E} \bigg|_{E=0}
\end{equation}
and magnetoelectric susceptibility 
\begin{equation}
\chi_{ME} =  \frac{\partial M}{\partial E} \bigg|_{E=0},
\end{equation}
which are also listed in Table~\ref{tab.E0Landau}. From the equilibrium solutions for $M$ and $P$ it is also straight forward to evaluate the free energy $F$, from which the entropy 
\begin{equation}
S = -\left( \frac{\partial F}{\partial T} \right)_E
\end{equation}
and specific heat 
\begin{equation}
C = -T\left( \frac{\partial^2 F}{\partial T^2} \right)_E
\end{equation}
can be evaluated. It is useful to note that $\frac{\partial a_M}{\partial T} = \alpha_M$ and $\frac{\partial a_P}{\partial T} = \alpha_P$.
 
\begin{table}[]
\caption{Solutions to the zero field Landau theory in Eq.~\eqref{eq.LandauF} (with $E=0$), as well as corresponding susceptibilities, free energy, entropy and specific heat. }
\begin{tabular}{l|c|c|c|c} 
\hline\hline
            & \multicolumn{1}{l|}{$M_0 = 0$, $P_0 = 0$} & \multicolumn{1}{l|}{$M_0 = 0$, $P_0^2 = \frac{-a_P}{b_P}$} & \multicolumn{1}{l|}{$M_0^2 = \frac{-a_M}{b_M}$, $P_0 = 0$} & \multicolumn{1}{c}{$M_0^2 = \frac{\lambda a_P - a_M b_P}{b_M b_P - \lambda^2}$, $P_0^2 = \frac{\lambda a_M - b_M a_P}{b_M b_P - \lambda^2}$} \\ \hline
$\chi_E$    & $a_P^{-1}$                                & $-\frac{1}{2}(a_P)^{-1}$                                 & $(a_P -\lambda \frac{a_M}{b_M})^{-1}$      			  & $-\frac{1}{2}(a_P -\lambda \frac{a_M}{b_M})^{-1}$ 	\\ \hline
$\chi_{ME}$ & 0                                         & 0                                                        & 0                                                        & $-\frac{\lambda P_0}{b_M M_0} \chi_E = -\frac{\lambda}{2}(\lambda^2 a_M a_P + a_M a_P b_M b_P - \lambda a_P^2 b_M - \lambda a_M^2 b_p)^{-1/2}$				\\ \hline
$F$         & 0                                         & $\frac{-a_P^2}{4b_P}$                                  & $\frac{-a_M^2}{4b_M}$                                  & $\frac{2 \lambda a_M a_P - a_P^2 b_M - a_M^2 b_P}{4(b_M b_P - \lambda^2)}$ \\ \hline
$S$         & 0                                         & $\frac{\alpha_P a_P}{2 b_P}$             & $\frac{\alpha_M a_M}{2 b_M}$             & -$\frac{( \lambda (\alpha_M a_P + a_M \alpha_P) - \alpha_P a_P b_M - \alpha_M a_M b_P)}{2(b_M b_P - \lambda^2)}$ \\ \hline
$C$         & 0                                         & $\frac{\alpha_P^2}{2 b_P}T$        				       & $\frac{\alpha_M^2}{2 b_M}T$   					          & -$\frac{( 2\lambda \alpha_M \alpha_P - \alpha_P^2 b_M - \alpha_M^2 b_P)}{2(b_M b_P - \lambda^2)}T$ \\ \hline\hline
\end{tabular}
\label{tab.E0Landau}
\end{table}

For $E \neq 0$, Eq.~\eqref{eq.Pcond} can no longer be solved as a quadratic for $P^2$ and it is more convenient to produce numerical solutions. After eliminating $M$, the free energy in the phase with $M \neq 0$ and $P \neq 0$ is 
\begin{equation}
F = -\frac{a_M^2}{4 b_M} + \frac{1}{2} \left( a_P - \frac{\lambda a_M}{ b_M}\right) P^2 + \frac{1}{4} \left( b_P - \frac{\lambda^2}{b_M} \right) P^4 - EP
\end{equation}
and the entropy at a fixed $P$ is 
\begin{equation}
S(T,E) = -\left( \frac{\partial F}{\partial T} \right)_E  = \frac{\alpha_M^2}{2b_M}(T- T_\mathrm{c}^M) - \frac{1}{2} \left( \alpha_P -\lambda\frac{\alpha_M}{b_M} \right) P^2(T,E) .
\end{equation}
The isothermal entropy change, resulting from varying a field from $E_1$ to $E_2$, is then the difference in the entropy for $P (T,E_2)$ and $P (T,E_1)$. 
\begin{equation}\label{eq.deltaS}
\Delta S = -\frac{1}{2}\alpha_P \left[ P^2 (T,E_2) - P^2(T,E_1)\right] -\frac{1}{2}\alpha_M \left[ M^2 (T,E_2) - M^2(T,E_1)\right] ,
\end{equation}
which provides an obvious decomposition of the total entropy change into an electric and a magnetic contribution.

\begin{equation}
\Delta S = -\frac{1}{2} \left( \alpha_P - \lambda \frac{\alpha_M}{b_M} \right) \left[ P^2 (T,E_2) - P^2(T,E_1)\right]  
\end{equation}
is an alternative formulation of Eq.~\ref{eq.deltaS}, valid when the field does not cause the magnetic phase to change. 

If one applied a magnetic field instead of an electric, the equivalent results as above could be obtained with $P$ exchanged for $M$ and $E$ exchanged for $B$, assuming that $M$ would be a ferromagnetic order parameter which couples to a magnetic field. 

\begin{figure}[hbt!]
	\centering
	\includegraphics[width=0.55\textwidth]{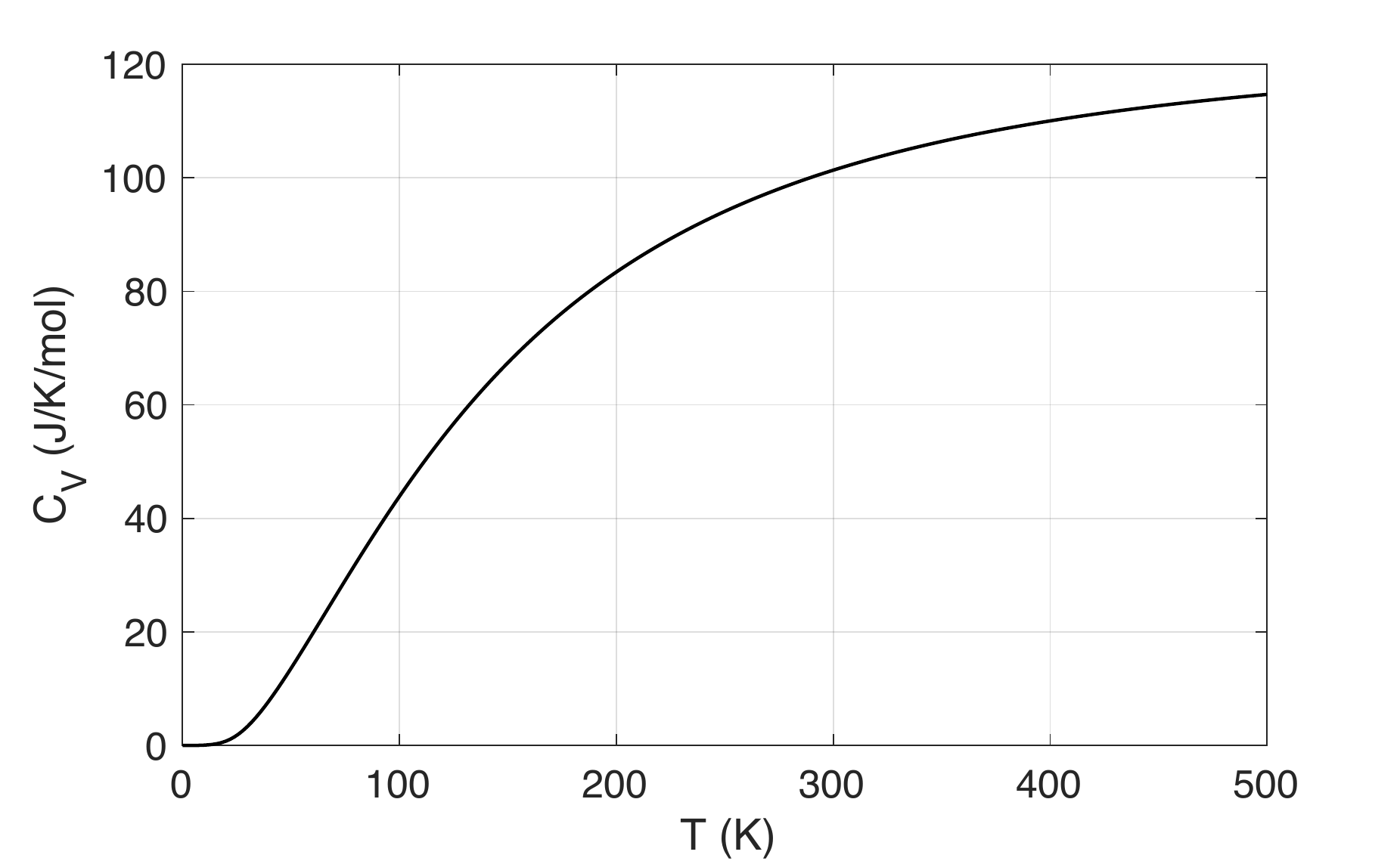}
	\caption{Phonon specific heat calculated from frozen phonon calculations. }
	\label{fig.Cv_of_T}
\end{figure}

From $\Delta S$, the adiabatic temperature change $\Delta T$ is estimated via $
\Delta T = -T \frac{\Delta S }{C}$, where $C$ is the specific heat. The temperature dependent specific heat is calculated from frozen phonon calculations for the cubic crystal structure, as shown in Fig.~\ref{fig.Cv_of_T}. These calculations largely agree with earlier such calculations which, however, neglected spin polarization~\cite{PhysRevB.75.214307}, whereas the result in Fig.~\ref{fig.Cv_of_T} was evaluated for G-type antiferromagnetism. 

%\appendix
%\section{Appendix title}\label{AppA}

\bibliography{literature}{}
\bibliographystyle{apsrev4-1}